\documentclass[amsmath, preprintnumbers, 10pt, twocolumn, pre bibliograpy, floatfix, superscriptaddress]{revtex4-1}

\usepackage{graphicx}
\usepackage{color}

\usepackage{physics}
\usepackage{bbm}
\usepackage{braket}
\usepackage[export]{adjustbox}
\usepackage{dsfont}
\usepackage{mathrsfs}
\usepackage{multirow}
\usepackage{float}
\usepackage{tabularx}
\usepackage{caption}
\usepackage[font=small,labelfont=bf,justification=RaggedRight,format=plain, singlelinecheck=false]{caption}
\usepackage{subcaption}
\usepackage{array}
\usepackage{hyperref}
\usepackage{cleveref}
\usepackage{amsmath}
\usepackage{amssymb}
\usepackage{stackengine}
\usepackage{mathrsfs}
\usepackage{ulem}

\renewcommand{\theequation}{\arabic{equation}}
\numberwithin{equation}{section}
\DeclareCaptionSubType*{figure}
\renewcommand\thesubfigure{\thefigure\alph{subfigure}}
\usepackage{xr}

\begin{document}
\title{Enhancing associative memory recall in non-equilibrium materials through activity}

\author{Agnish Kumar Behera}
\affiliation{Department of Chemistry, University of Chicago, Chicago, IL, 60637}
\author{Madan Rao}
\affiliation{Simons Centre for the Study of Living Machines, National Centre for Biological Sciences - TIFR, Bangalore}
\author{Srikanth Sastry}
\affiliation{Jawaharlal Nehru Centre For Advanced Scientific Research, Bangalore}
\author{Suriyanarayanan Vaikuntanathan*}
\affiliation{Department of Chemistry, University of Chicago, Chicago, IL, 60637}
\affiliation{The James Franck Institute, University of Chicago, Chicago, IL, 60637}

\setlength{\abovedisplayskip}{2pt}
\setlength{\belowdisplayskip}{2pt}
\setlength{\abovedisplayshortskip}{2pt}
\setlength{\belowdisplayshortskip}{2pt}

\begin{abstract}
Associative memory, a form of content-addressable memory, facilitates information storage and retrieval in  many biological and physical systems. In statistical mechanics models, associative memory at equilibrium is represented through attractor basins in the free energy landscape. Here, we use the Hopfield model, a paradigmatic model to describe associate memory, to  investigate the effect of non-equilibrium  activity on memory retention and recall. We introduce activity into the system as gaussian-colored noise which breaks detailed balance and forces the system out of equilibrium. We observe that, under these non-equilibrium conditions, the Hopfield network has a higher storage capacity than that allowed at equilibrium. Using  analytical and numerical techniques, we show that the rate of entropy production modifies the energy landscape and helps the system to access memory regions which were previously inaccessible.
\end{abstract}
\maketitle

\renewcommand*{\thesection}{\Roman{section}}
\section{Introduction}
\renewcommand*{\thesection}{\arabic{section}}

Biological systems ranging from neuronal circuits in multicellular organisms, to biological circuits responsible for immune memory, display a remarkable array of information storage and retrieval dynamics\cite{Identi1999, MemoryAndBrain, Peretto1984, Little1996, Barton1965}. Understanding the statistical mechanical basis of memory and computation in such systems is crucial, particularly in the context of efforts seeking to recreate such phenomenology in artificial systems~\cite{Hebb, Hopfield2554, Hopfield1984, MemoryReview}. Statistical mechanical models, such as the paradigmatic Hopfield model, have for instance been used to show how the capacity of a system for associative memory may be explained in terms of the underlying free energy landscapes \cite{Hopfield2554, Crisant_statics}. While these studies have established limits on the capacity of a system to store and retrieve information, most of these explanations are based on analysing systems evolving according to equilibrium or near-equilibrium dynamics\cite{Crisant_statics, Crisanti_dynamics}. In this paper, we consider how non-equilibrium forcing and dissipation may modify the associative memory capacity of a system. Our main results suggest that non-equilibrium driving may indeed enhance the associative memory properties of a system and make it more robust. \\
We demonstrate our results using a version of the Hopfield model \cite{Bolle2003}. The Hopfield model is one of the most widely studied systems for understanding associative memory and shows how a collection of Ising like spins can be used to store multiple patterns or memories. Seminal calculations using equilibrium statistical mechanics have shown how the patterns or memories are encoded as local minima in the energy landscapes of the spin system \cite{Sompolinsky1985, amit_1989, Crisant_statics}. Importantly, these models exhibit tradeoffs between the amount of memory stored and the ability to robustly retrieve memories. Under equilibrium dynamics, these tradeoffs imply that only a finite amount of memories or patterns maybe a stored in an associative memory system. Attempting to store patterns beyond this capacity leads to a degradation in the memory retrieval properties of the system\cite{McEliece1987, amit_1989}. 
Non-equilibrium dynamics and dissipation may potentially provide an opportunity to improve the associative memory capacity in such systems. Indeed, non-equilibrium activity has been used to stabilize assembly and promote the formation of ordered states in a variety of active and driven systems \cite{Redner2016, MNguyen2021, ACaccuito2018, MadanRSC, Ramaswamy2019}. In some limited contexts, the stabilizing role of non-equilibrium activity maybe understood in terms of deeper or new basins of attraction surrounding ordered states in  an effective free energy landscape\cite{MECates2015, Etienne2016, Maitra2020}. These connections between non-equilibrium forcing and the promotion of specific ordered states raise the intriguing prospect that non-equilibrium activity can potentially be used to enhance memory capacity and retrieval in associative memory systems. Here, by endowing a version of the Hopfield model with non-equilibrium dynamics, we demonstrate how the associative memory characteristics of a system may indeed be enhanced. \\
The rest of the paper is organized as follows. In Section \ref{ModelAndSimulations}, we describe the main associative memory model considered in this paper, namely, a so called \textit{spherical} Hopfield like model consisting of spins that can take on a range of continuous values (as opposed to the discrete Ising like spins in the canonical version of the Hopfield model). Motivated by phenomenology seen in minimal active Brownian particle systems, the dynamics of the spins in this spherical Hopfield model are driven away from equilibrium by subjecting them to detailed balance violating colored noise \cite{Etienne2016}. The temporal correlation in the colored noise do have an equivalent counterpart in the temporal correlations of the drag force, thus our system is out of equilibrium. Using numerical simulations, we show how these detailed balance violating dynamics surprisingly succeed in improving memory recall. Next, in Section \ref{AnalyticalCalculations}, we provide a variety of analytical arguments to rationalize and explain these observations. First, in a perturbative limit that allows for the description of the non-equilibrium configurational steady state in terms of an effective free energy landscape, we show how the effect of non-equilibrium driving leads to the formation of more stable free energy basins around stored patterns\cite{MECates2015, Maitra2020}. Next, using the Martin-Siggia-Rose generating functional formalism to move beyond the perturbative limits, we again show how the addition of activity leads to improved memory recall\cite{MSR1973, COOLEN2001619, Bolle2003}. Finally, in Section \ref{MSR_main_section} and \ref{Conclusions}, we use these analytical results to illustrate how activity improves associative memory in a broad range of parameters. Together, our results suggest new and general ways in which non-equilibrium forcing and dissipation maybe used to enhance the information processing abilities of a material. 

\begin{figure}[thb]
    \begin{subfigure}[b]{0.39\textwidth}
    \centering
    \includegraphics[width = \textwidth]{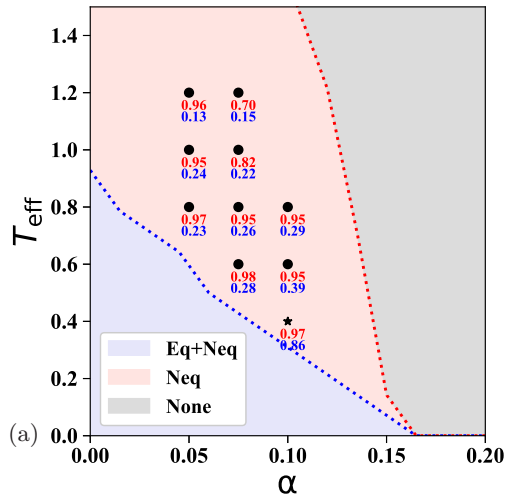}\llap{\parbox[b]{5.2in}{(a)\\\rule{0ex}{0.35in}}}
    \captionlistentry{}
    \label{N200NeqEqComparison}
    \end{subfigure}
    \begin{subfigure}[b]{0.4\textwidth}
    \centering
    \includegraphics[width = \textwidth]{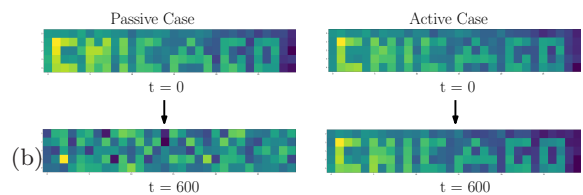}\llap{\parbox[b]{5.8in}{(b)\\\rule{0ex}{0.1in}}}
    \captionlistentry{}
    \label{ChicagoPattern}
    \end{subfigure}
    \caption{ Memory retrieval in an associative memory model with equilibrium and non-equilibrium dynamics (a) A phase diagram demarcating regions exhibiting associative memory. The \textit{blue} region represents the parameter space where retrieval is possible in both the passive (equilibrium) and  active (out of equilibrium) case, the \textit{red} region is where the active case shows retrieval whereas the passive system does not and in the \textit{grey} region memory retrieval is lost altogether. The phase boundaries were obtained using a mean field technique described in SI Sec. \ref{MSRSupp}. The bold circles mark the values of effective temperature and $\alpha$ at which numerical simulations using  Eq.~\ref{RelaxationalLangevin} were performed. The numbers in red denote the final overlap parameter (Eq.~\ref{OrderParameter1}) for ``condensed patterns" in the active case, those in blue denote the value in the passive case. All the non-equilibrium dynamics shown here were performed with $\tau=5.0$. (b) As an example, we show the retrieval dynamics of a pattern stored  at the location \textit{``*"} in the phase diagram above i.e. at $\alpha=0.4$, $T_{\rm eff} = 0.4$. For illustrative purposes, the pattern is arranged such that it spells out ``Chicago". Simulations with equilibrium dynamics, when initialized in the vicinity of this pattern, fail to retrieve it. On the other hand, this pattern is successfully retrieved with non-equilibrium dynamics at the same effective temperature. These numerical simulations were performed with $N=200$ spins with $15$ patterns encoded in the interactions.}
    \label{Fig2}
\end{figure}

\renewcommand*{\thesection}{\Roman{section}}
\section{A Spherical Hopfield model with non-equilibrium dynamics}
\label{ModelAndSimulations}
\renewcommand*{\thesection}{\arabic{section}}
The Hopfield model is an interacting spin system with a Hamiltonian which is fully connected, i.e. every spin is connected to every other spin. We work with a version of the Hopfield model \cite{Bolle2003} where the spins are continuous and obey the constraint, $\sum_{i=1}^N \sigma_i^2 = N$, where $\sigma 's$ denote the spins and N is the total number of spins in the system.
\begin{align}
    \mathcal{H}_0 &= \frac{1}{2} \mu \sigma_i \sigma_i - \frac{1}{2}J_{ij} \sigma_i \sigma_j - \frac{u_0}{4}J_{ijkl} \sigma_i \sigma_j \sigma_k \sigma_l \label{HopfieldHamiltonian} \\
    J_{ij} &= \frac{1}{N} \xi^{\mu}_i \xi^{\mu}_j \ , \ J_{ijkl} = \frac{1}{N^3} \xi^{\mu}_i \xi^{\mu}_j \xi^{\mu}_k \xi^{\mu}_l \label{Jij_Jijkl} 
\end{align}

Here repeated indices have been summed over. The spin variables in the system are denoted by $\sigma_i$, where $i$ denotes the site index and the pattern variables as $\xi^{\mu}_i$ where $\mu$ denotes the pattern index and $i$ denotes the site index. The N components of the patterns are drawn from i.i.d. normal distributions, $\xi_i^{\mu} \sim \mathscr{N}(0,1)$. In the model, the coupling strengths between spins, $J_{ij}$ and $J_{ijkl}$  depend on the patterns through the Hebbian rule \cite{Hebb}. The Hebbian Rule states that ``neurons which fire together, wire together". When two spins in a given pattern have the same sign, the connection strength between them is increased and vice versa. The network stores the patterns by modifying the weights (connection strength) between the spins. Quartic terms are included in the Hamiltonian following Ref~\cite{Bolle2003} where it was demonstrated that such higher order terms are a necessary requirement for associative memory like properties in a system with continuous spins. 
The spins evolve according to the following equations of motion,
\begin{align}
    \partial_t \sigma_i = -\mu(t) \sigma_i(t)  - \frac{\delta \mathcal{H}_0(\mathbf{\sigma})}{\delta \sigma_i (t)} + \eta_i(t) \label{RelaxationalLangevin}
\end{align}
Here, $\mu$ is the Lagrange multiplier which ensures the normalization of the spins and $\frac{\delta \mathcal{H}_0}{\delta \sigma}$ is the relaxational term  Finally, $\eta(t)$ models the effect of various thermal and athermal fluctuations, 
\begin{align}
    \eta_i(t) &= \eta_{w,i}(t) + \eta_{a,i}(t) \label{NoiseTotal}\\
    \braket{\eta_{w,i}(t)} &= 0 = \braket{\eta_{a,i}(t)} \ \forall \ i, \ t \label{NoiseMean}\\
    \braket{\eta_{w, i}(t) \eta_{w, j}(t')} &= 2T_p \delta_{ij} \delta(t-t') \label{WhiteCorr}\\
    \braket{\eta_{a,i}(t) \eta_{a,j}(t')} &= \frac{T_a}{\tau} \delta_{ij} \exp\left(-\frac{|t - t'|}{\tau} \right) \label{ActiveCorr}
\end{align}
where the thermal fluctuations (thermal noise) are modelled using a delta function correlated white noise, $\vec{\eta_w}$, and the exponentially correlated  $\vec{\eta_a}$, is a so called colored noise source. 
As mentioned earlier, the addition of colored noise into the system without a corresponding change in dissipation breaks detailed balance as has been demonstrated in Refs.\,\cite{Bonilla2019, Kubo1966, Etienne2016}. Thus even in the limit $\tau \to 0$, this will still be nonequilibrium. For this system, we define an \text{effective temperature} as $T_{\rm eff} = T_p + T_a$. While our non-equilibrium system isn't generically described by an effective temperature \textendash indeed, the  effective temperature $T_{\rm eff}$ is most appropriate in the low persistence time limit \cite{Maitra2020, Fodor_2020} \textendash $T_{\rm eff}$ provides a convenient way to characterize the strength of the non-equilibrium forcing. 
\begin{align}
    \lim_{\tau \to 0} \braket{\eta_{a,i}(t) \eta_{a,j}(t')} &= 2 T_a \delta(t - t')
\end{align}

We perform numerical simulations in which a system with $N=200$ spins is evolved forward in time using Eq.~\ref{RelaxationalLangevin}. In these numerical simulations, we probe the ability of the system to retrieve a stored pattern by initializing the spin system in configurations close to those corresponding to the stored memory states. Retrieval is considered successful if the dynamics are able to recover the full stored pattern when the system reaches an approximate steady state characterized by an almost constant overlap value. 
Quantitatively, the retrieval ability is measured by tracking the steady state value of the overlap of the final spin configuration of the system with the pattern it was initialized near. For a particular pattern $\mu$, this overlap can be measured as $\frac{1}{N} \sum_i \xi^{\mu}_i \sigma_i$. In Fig. \ref{N200NeqEqComparison}, we describe the retrieval phase diagram for this system as a function of $T_{\rm eff}$ and $\alpha$, where $\alpha$ determines the number of patterns encoded in the Hamiltonian according to $\alpha N$, for both equilibrium  and non-equilibrium dynamics. Here, for equilibrium dynamics, $T_{\rm eff}$ is simply equal to the passive temperature, $T_p$. The non-equilibrium dynamics chosen for the simulations in Fig.~\ref{ChicagoPattern}(a) were performed with $T_{\rm eff}=0.4\,,\tau=5$. Qualitatively similar results can be obtained for other choices of the non-equilibrium parameters. As can be clearly seen in Fig.~\ref{N200NeqEqComparison}, the ability of the spin system to retrieve patterns is markedly increased due to non-equilibrium driving. Specifically, the blue shaded region demarcates the parameter combinations under which associative memory or memory retrieval is possible under equilibrium dynamics~\cite{Sompolinsky1985}. Our non-equilibrium simulations demonstrate memory retrieval in the red shaded region in addition to the blue shaded region. This can be observed through the numbers in \textit{red} (active) which are consistently higher than those in \textit{blue} (passive) in the \textit{red} region of the phase plot. Thus, given the same $T_{\rm eff}$, these numerical results show how non-equilibrium forcing allows the spin system to store larger number of patterns. 

In the subsequent sections, we explore the theoretical basis of this improved associative memory due to non-equilibrium dynamics. First in Section \ref{RepCal} we perform a perturbative analysis in the limit of small persistence time, $\tau$. In this limit the nonequilibrium distribution function can be approximated using Boltzmann statistics with an effective Hamiltonian and an effective temperature. Our calculations show how the effective Hamiltonian supports enhanced interactions between spins as well as new higher order interactions at first order in $\tau$. A replica calculation reveals that as a consequence of these enhanced interactions \textendash these emerge due to the non-equilibrium forcing \textendash the spin system possesses enhanced associative memory recall.  Then in Section \ref{MSR_main_section}, we perform a Martin-Siggia-Rose calculation which describes our system in a mean field limit and provides an analytically tractable route to quantify how the robustness of pattern retrieval increases away from equilibrium. \\

\renewcommand*{\thesection}{\Roman{section}}
\section{Rationalizing improved associative memory under non-equilibrium dynamics }
\label{AnalyticalCalculations}
\renewcommand*{\thesection}{\arabic{section}}

The equilibrium Hopfield model can be solved analytically using the replica method \cite{Sompolinsky1985}. Since our model is out of equilibrium, a direct application of the replica method is not possible. Through Unified Active Noise Approximation (UCNA) \cite{Hanngi1987} and recent work by authors of Ref. \cite{Maitra2020} we show that our active system can be described using an \textit{effective} Hamiltonian and a new \textit{effective} temperature. We then use the standard replica technique with this effective Hamiltonian to show how the addition of activity enhances associative memory recall. In a subsequent section \ref{MSR_main_section} we derive an exact mean field set of evolution equations for our active system using the Martin-Siggia-Rose generating functional formalism \cite{MSR1973,COOLEN2001619, Bolle2003} and further illustrate how memory recall is improved by the introduction of activity. 
\renewcommand*{\thesection}{\Roman{section}}
\subsection{Effective interactions due to non-equilibrium forcing provide a mechanism for improved associative memory recall}
\label{RepCal}
\renewcommand*{\thesection}{\arabic{section}}

Unified Colored Noise approximation (UCNA) \cite{Hanngi1987, Etienne2016} suggests that at small $\tau$ our non-equilibrium system can be described by an effective Hamiltonian with an effective temperature. 
As outlined in SI Sec.~\ref{EffectiveHamiltonianSupp}, we show that at first order in $\tau$, the perturbed Hamiltonian and the effective temperature ($\tilde{T}$) are given by,
\begin{align}
    \mathcal{H} &= \mathcal{H}_0 + \frac{\tau T_a}{\tilde{T}}\left(\frac{1}{2} |\nabla_{\sigma} \mathcal{H}_0|^2 - \nabla^2_{\sigma} \mathcal{H}_0 \right) \label{PerturbedHamiltonian} \\
    \tilde{T} &= T_p + T_a \label{Effective_Temperature}
\end{align}
Substituting the Hopfield Hamiltonian, Eq.~\ref{HopfieldHamiltonian} into Eq.~\ref{PerturbedHamiltonian} yields,
\begin{align}
    \mathcal{H} =& \frac{1}{2} \mu \sigma_i^2 - \frac{v}{2}J_{ij} \sigma_i \sigma_j - \frac{u}{4}J_{ijkl} \sigma_i \sigma_j \sigma_k \sigma_l \nonumber \\
    &+ \frac{k}{6}J_{ijklmn} \sigma_i \sigma_j \sigma_k \sigma_l \sigma_m \sigma_n + O(1/N) \ terms \label{EffectiveHamiltonian} \\
    v =& 1 + \frac{\tau T_a(2\mu - s)}{\tilde{T}} ,\ u = u\left[1 + \frac{2\tau T_a(2\mu - 2s)}{\tilde{T}} \right] \\
    k =& \frac{3\tau T_a u^2 s}{\tilde{T}} , \ s = \braket{(\xi^{\mu}_i)^2} \ , \ J_{ijklmn} = \frac{\xi^{\mu}_i \xi^{\mu}_j \xi^{\mu}_k \xi^{\mu}_l \xi^{\mu}_m \xi^{\mu}_n}{N^5}  
\end{align}

\begin{figure}[th]
    \begin{subfigure}[t]{0.4\textwidth}
    \centering
    \includegraphics[width =
    \textwidth]{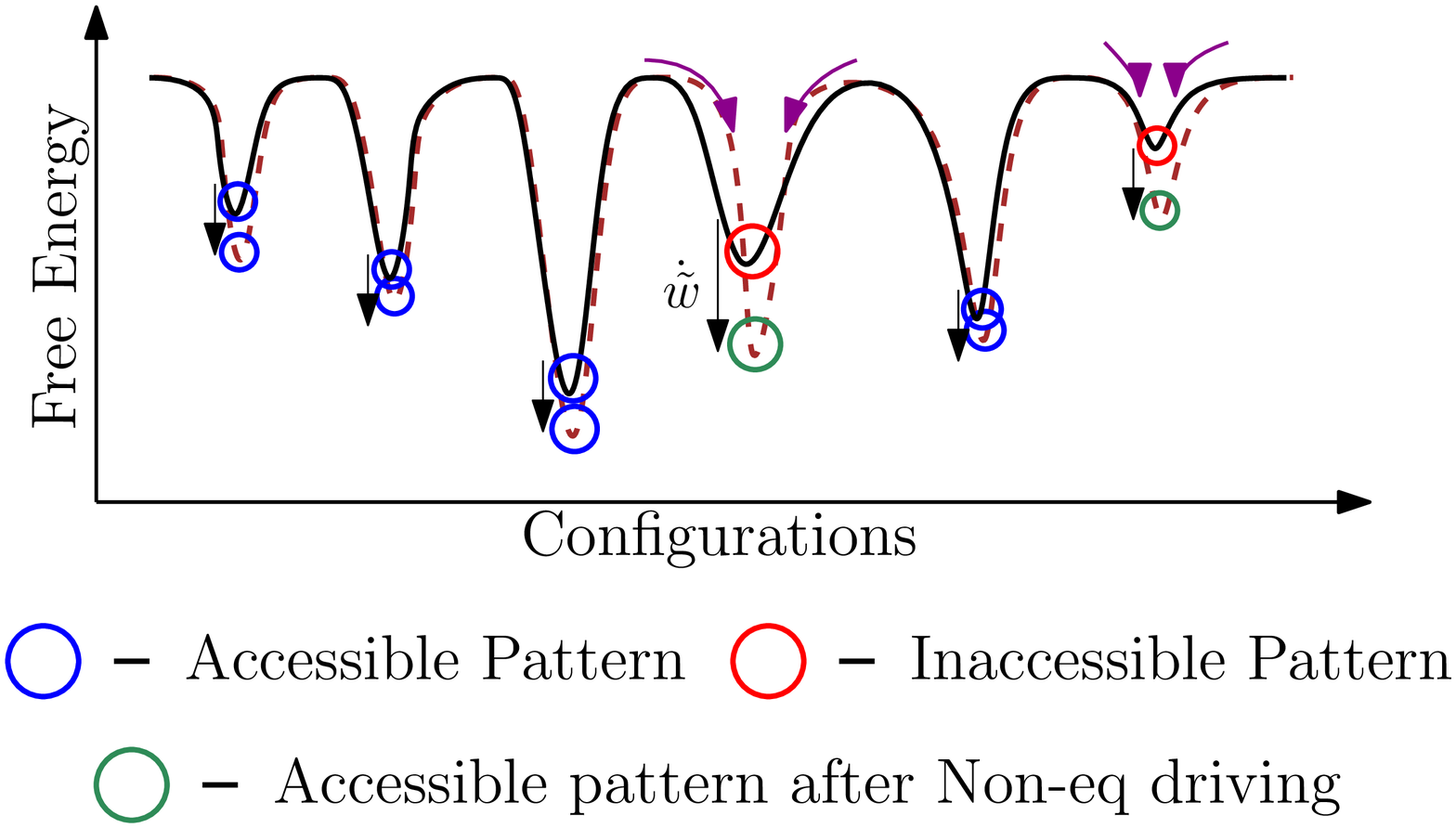}\llap{\parbox[b]{5.5in}{(a)\\\rule{0ex}{1.5in}}}
    \captionlistentry{}
    \label{FreeEnergyLandscape}
    \end{subfigure}
    \begin{subfigure}[t]{0.4\textwidth}
    \centering
    \includegraphics[width = \textwidth]{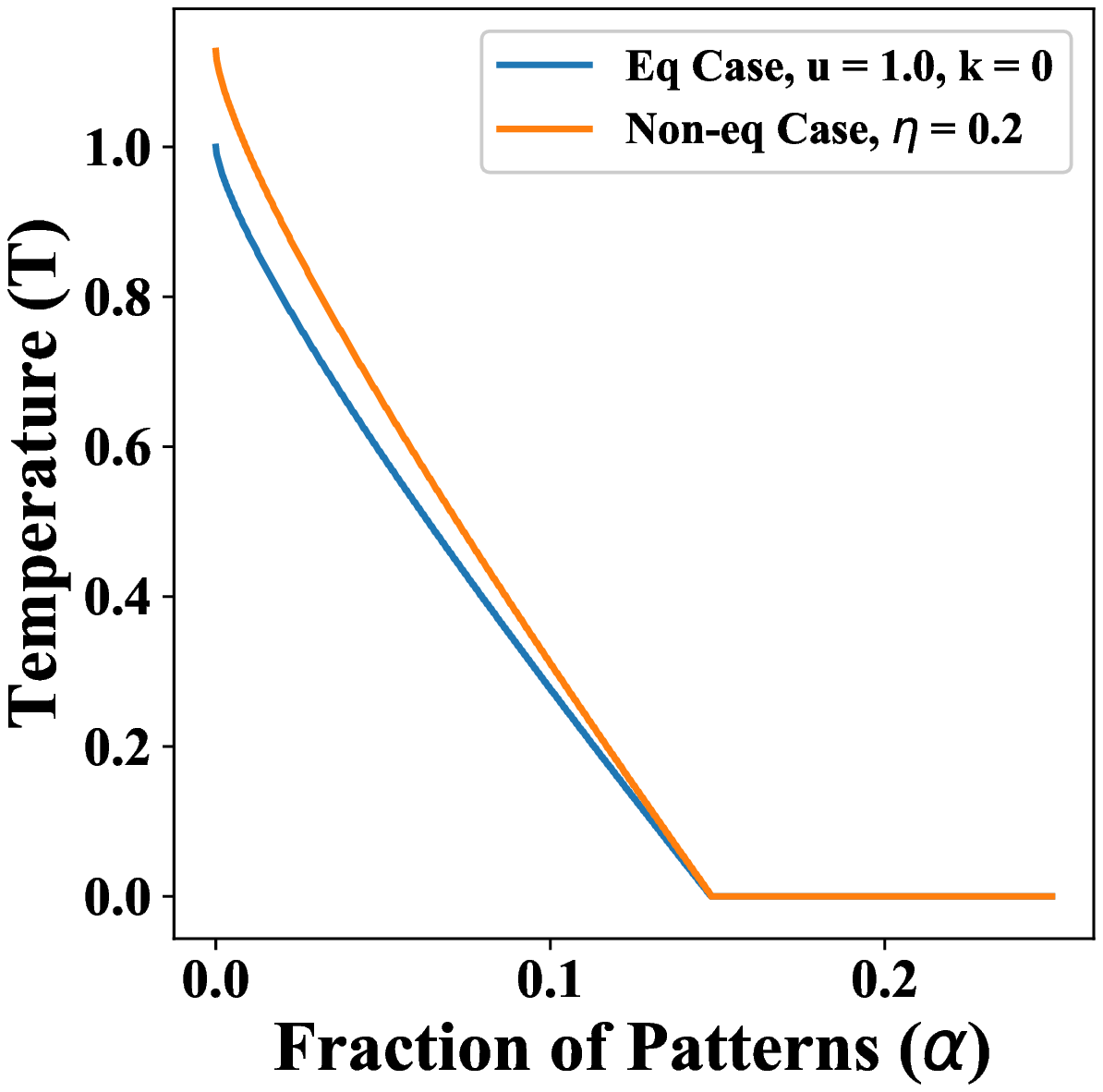}\llap{\parbox[b]{4.9in}{(b)\\\rule{0ex}{2.6in}}}\llap{\parbox[b]{2in}{No Memory\\\rule{0ex}{1.7in}}}\llap{\parbox[b]{3.8in}{Memory\\\rule{0ex}{0.8in}}}
    \captionlistentry{}
    \label{NeqEqComparison}
    \end{subfigure}
    \caption{Enhancement of associative memory recall due to non-equilibrium activity. (a) A schematic of how the free energy landscape gets modified due to non-equilibrium driving. In the region where the system is able to retrieve patterns, nonequilibrium driving leads to deepening of the energy basins near the patterns. Thus patterns which could not be retrieved in the equilibrium phase can be retrieved in the presence of activity.
    (b) We demarcate the boundary between regimes with associative memory and those without by using the Replica Calculation.  The region to the left of the lines labelled using ``Memory" represents the retrieval phase. With a passive fraction, $f = 0.2$, $\tau = 0.5$, $s = 0.8$, and Lagrange Multiplier $\mu = 2$, and taking into account the first-order corrections, the phase diagram shows an enhancement in memory, i.e. the overlap parameter $m$ is non-zero over a larger region.}
\end{figure}

Thus we see that activity enhances the strength of the quadratic and quartic terms. It also generates higher order coupling between spins, for instance at first order it gives rise to a sextic term. We note that the extra terms due to activity maybe related to the dissipative cost accompanying non-equilibrium forcing. Indeed, as we show in SI Sec.~\ref{RateOfWork_supp}, the correction to the interactions in Eq.~\ref{PerturbedHamiltonian} are on average equal to 
\begin{equation}
    \dot{\tilde{w}} = \langle \eta_a \cdot (\nabla_{\sigma} \mathcal{H} - \mu \sigma) \rangle - \langle \mu \sigma \cdot \nabla_{\sigma} \mathcal{H} \rangle\,,
    \label{eq:dotwDef}
\end{equation}
where  $\dot{\tilde{w}}$ is the rate of entropy production in the system and characterizes the dissipative cost associated with the non-equilibrium forcing.
This above can be derived by studying the rate of change of the potential function as outlined in Ref.~\cite{Laura2019}. The enhanced strength of the quadratic and quartic terms in Eq.~\ref{PerturbedHamiltonian} may potentially offer a route to stabilize associative memory recall by modifying the effective free energy landscape supporting pattern retrieval (Fig.~\ref{FreeEnergyLandscape}). We now confirm this picture by performing a  replica calculation on this effective Hamiltonian. Our calculation identifies regimes where the non-equilibrium model can exhibit associative memory whereas an equivalent equilibrium model fails to have associative memory properties. The details of the calculation are provided in SI Sec.~\ref{ReplicaCalculationSupp} and follow standard calculations performed on the equilibrium Hopfield model~\cite{amit_1989}.

The replica calculation provides an estimate of the order parameter $m$ that characterizes the degree of polarization of the system towards one of the stored patterns. 
\begin{equation}
    m = \frac{1}{N} \xi^{\mu}_i \sigma_i \label{OrderParameter1} \\
\end{equation}
A non-zero value of this order parameter, along with conditions on an additional order parameter that characterizes the spin glass nature of the system, identify regimes in which memory retrieval is possible. In Fig. \ref{NeqEqComparison} we plot the results from our replica calculation and demarcate regimes in which memory retrieval is possible both in and out of equilibrium. The specific calculation in Fig. \ref{NeqEqComparison} was performed with $T_p$ and $T_a$ fixed such that $ \frac{T_p}{T_{\rm eff}} = 0.2$. 
Non-equilibrium dynamics permit memory retrieval even in regions with higher effective temperatures. In general, the presence of higher order terms might lead to the formation of spurious energy minimas but even then the system can robustly retrieve the original patterns stored in it.

While, this is only an approximate result as the perturbation is valid only for small $\tau$ and we ignore the higher order corrections to the effective Hamiltonian, our results nonetheless show how the strengthening of stabilizing interactions in the effective Hamiltonian due to non-equilibrium activity results in improved associative memory recall.

\renewcommand*{\thesection}{\Roman{section}}
\subsection{A mean field approach using the Martin-Siggia-Rose Generating Functional  Lagrangian to explain improvement in non-equilibrium associative memory recall}
\label{MSR_main_section}
\renewcommand*{\thesection}{\arabic{section}}

The results of the previous section analytically show how associative memory can be improved in the low persistence time limit. In order to probe the effects of activity in other limits, we use the Martin-Siggia-Rose (MSR) generating functional approach to write down and study mean field coarse-grained equations for the evolution of the order parameter $m(t)$.

In the MSR approach, the statistics of the evolution path of the system are captured by the disorder averaged generating functional Eq.~\ref{GeneratingFunctional}.
\begin{multline}
    \overline{Z[\mathbf{\psi}]} = \int D\sigma D\eta P(\eta) \overline{\exp \left[ i \sum_{i=1}^N \int dt \psi_i(t) \sigma_i(t) \right] }  \\
    \overline{\prod_{i = 1}^{N} \delta \left(\partial_t \sigma_i(t) + \mu(t)\sigma_i(t) + \frac{\delta \mathcal{H}(\mathbf{\sigma})}{\delta \sigma_i(t)} - \eta_i(t) \right)} \label{GeneratingFunctional}
\end{multline}

When we perform the disorder average over $Z$ in Eq.~\ref{GeneratingFunctional} we decouple the spins but couple different times. This is similar to the coupling of different replicas in the Replica approach. The entire procedure is described in SI Sec.~\ref{MSRSupp}. We obtain an equation of motion for the decoupled spins following the procedure. Using the decoupled equation of motion for single spins, SI Eq.~\ref{DecoupledEqn}, we can easily write down the equations of motion for the macroscopic variable, m (SI Eq.\ref{EqnForm_analytical}).

\begin{widetext}

\begin{figure}[thb]
    \begin{subfigure}[t]{0.3\textwidth}
    \centering
    \includegraphics[width = \textwidth]{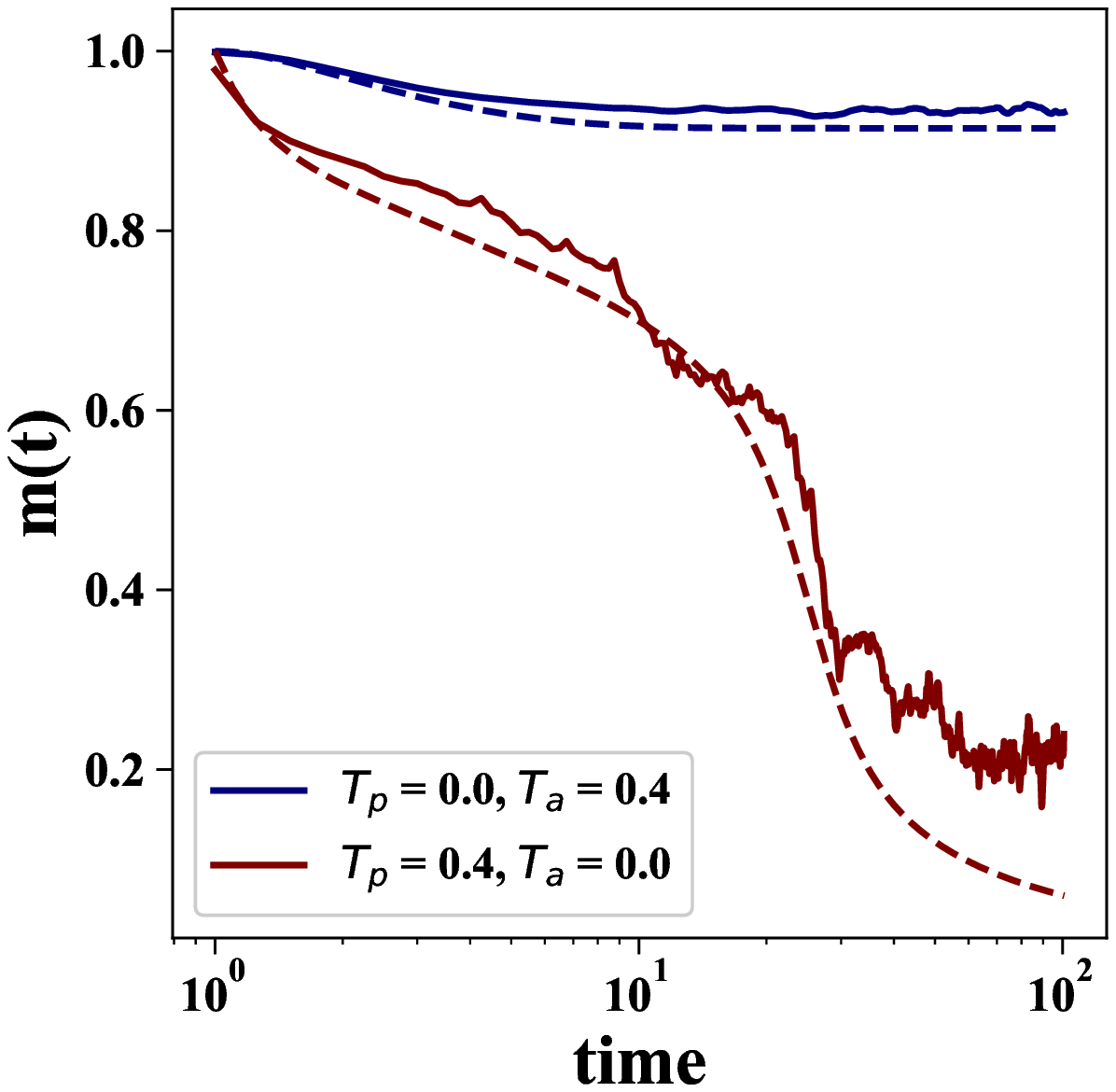}\llap{\parbox[b]{3.8in}{(a)\\\rule{0ex}{2in}}}
    \captionlistentry{}
    \label{GLE_SHF_Comparison}
    \end{subfigure}
    \begin{subfigure}[t]{0.3\textwidth}
    \centering
    \includegraphics[width = \textwidth]{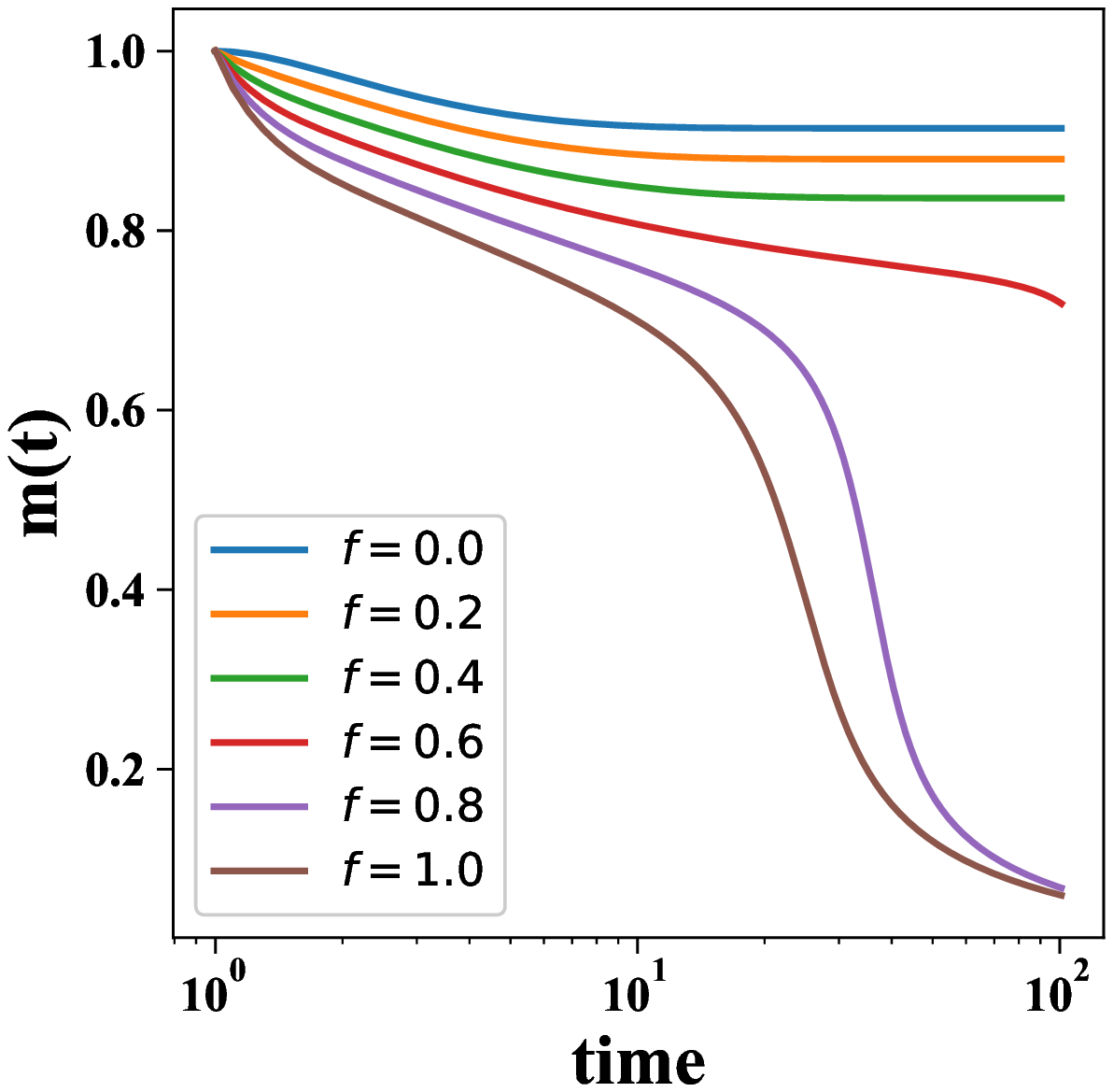}\llap{\parbox[b]{3.8in}{(b)\\\rule{0ex}{2in}}}
    \captionlistentry{}
    \label{mEvolution}
    \end{subfigure}
    \begin{subfigure}[t]{0.3\textwidth}
    \centering
    \includegraphics[width = \textwidth]{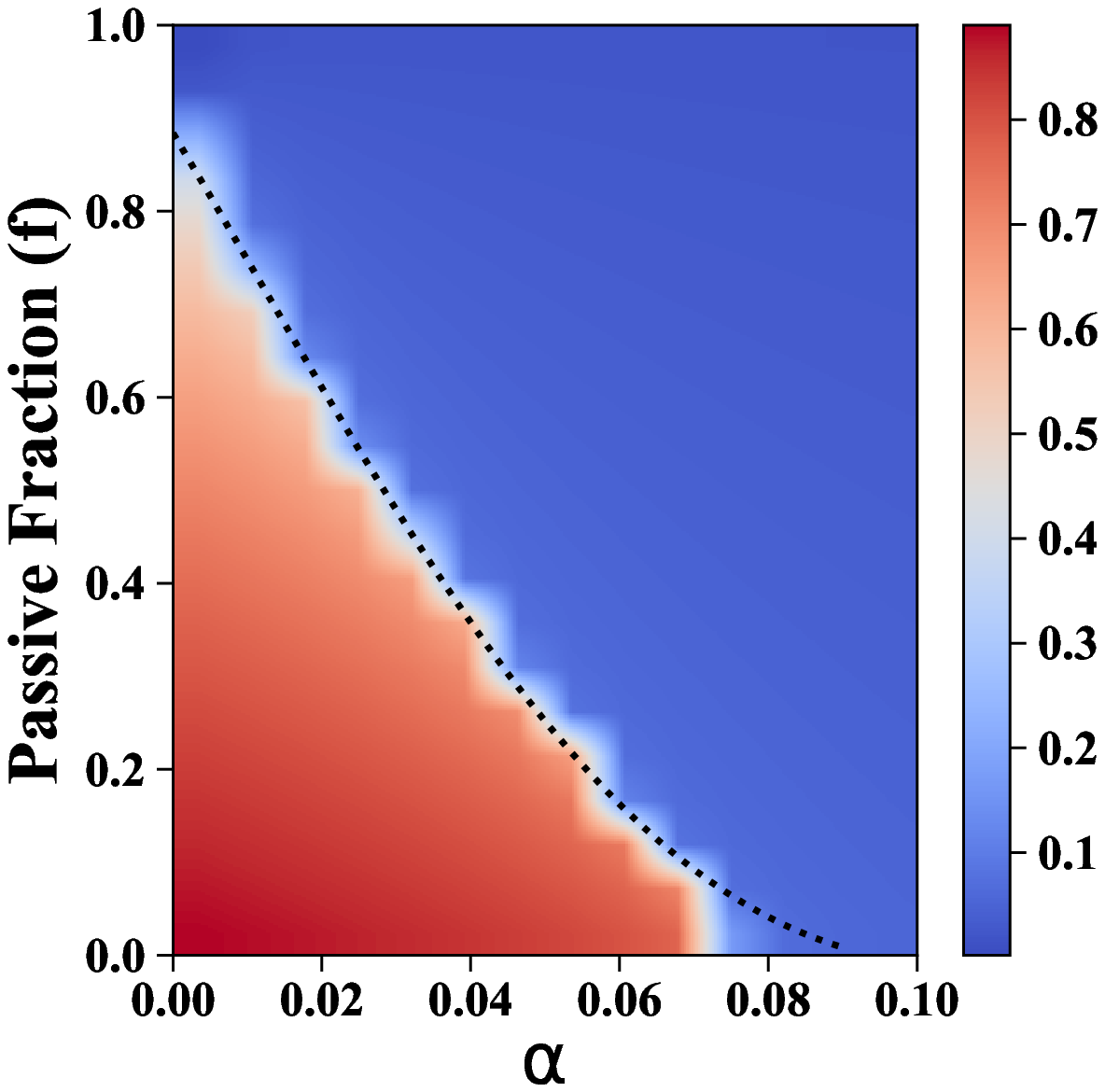}\llap{\parbox[b]{3.8in}{(c)\\\rule{0ex}{2in}}}
    \captionlistentry{}
    \label{NeqPhsPLot_tau_Teff_constant}
    \end{subfigure}
    \begin{subfigure}[t]{0.3\textwidth}
    \centering
    \includegraphics[width = \textwidth]{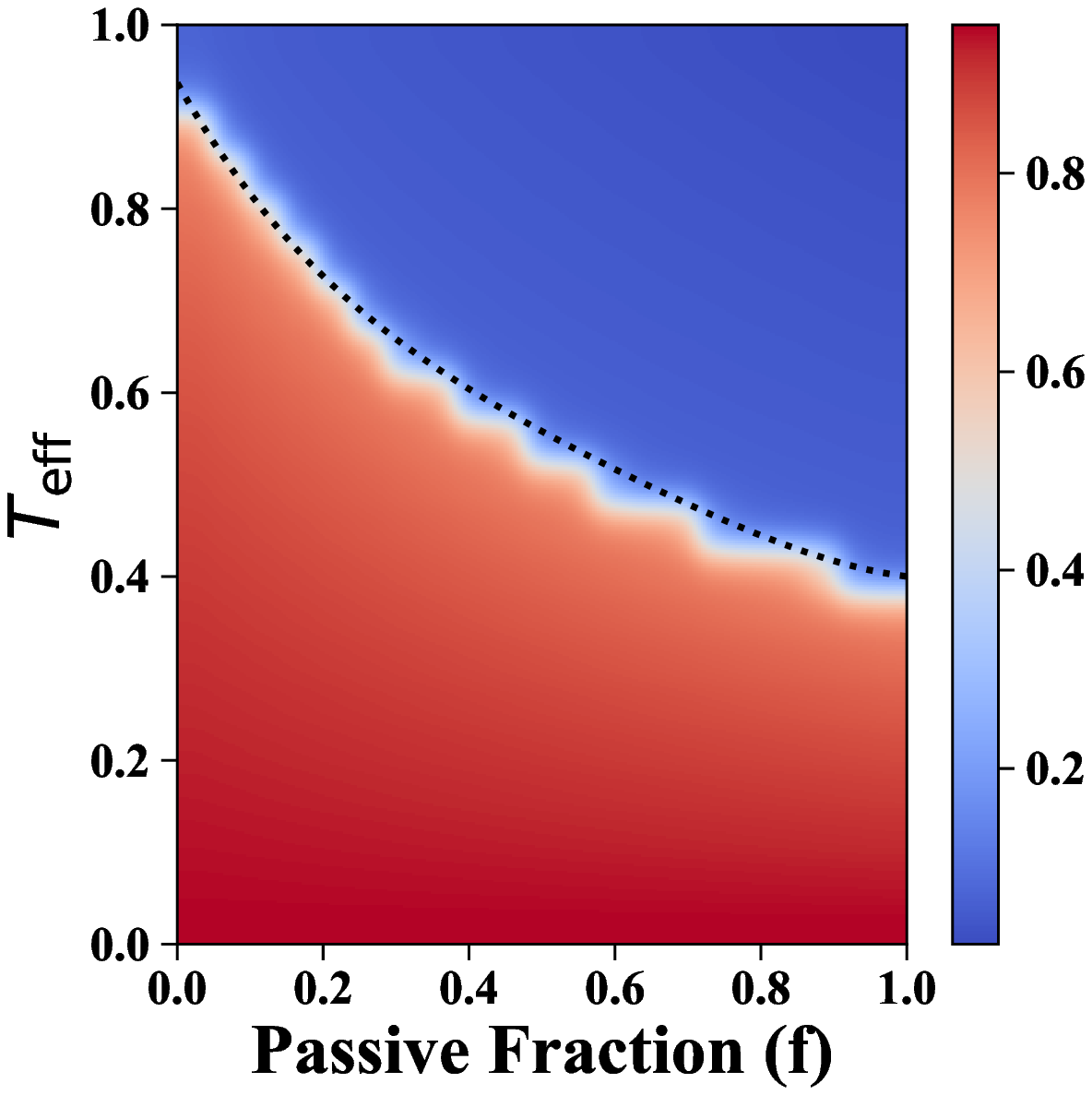}\llap{\parbox[b]{3.8in}{(d)\\\rule{0ex}{2in}}}
    \captionlistentry{}
    \label{NeqPhsPLot_tau_alpha_constant}
    \end{subfigure}
    \begin{subfigure}[t]{0.3\textwidth}
    \centering
    \includegraphics[width = \textwidth]{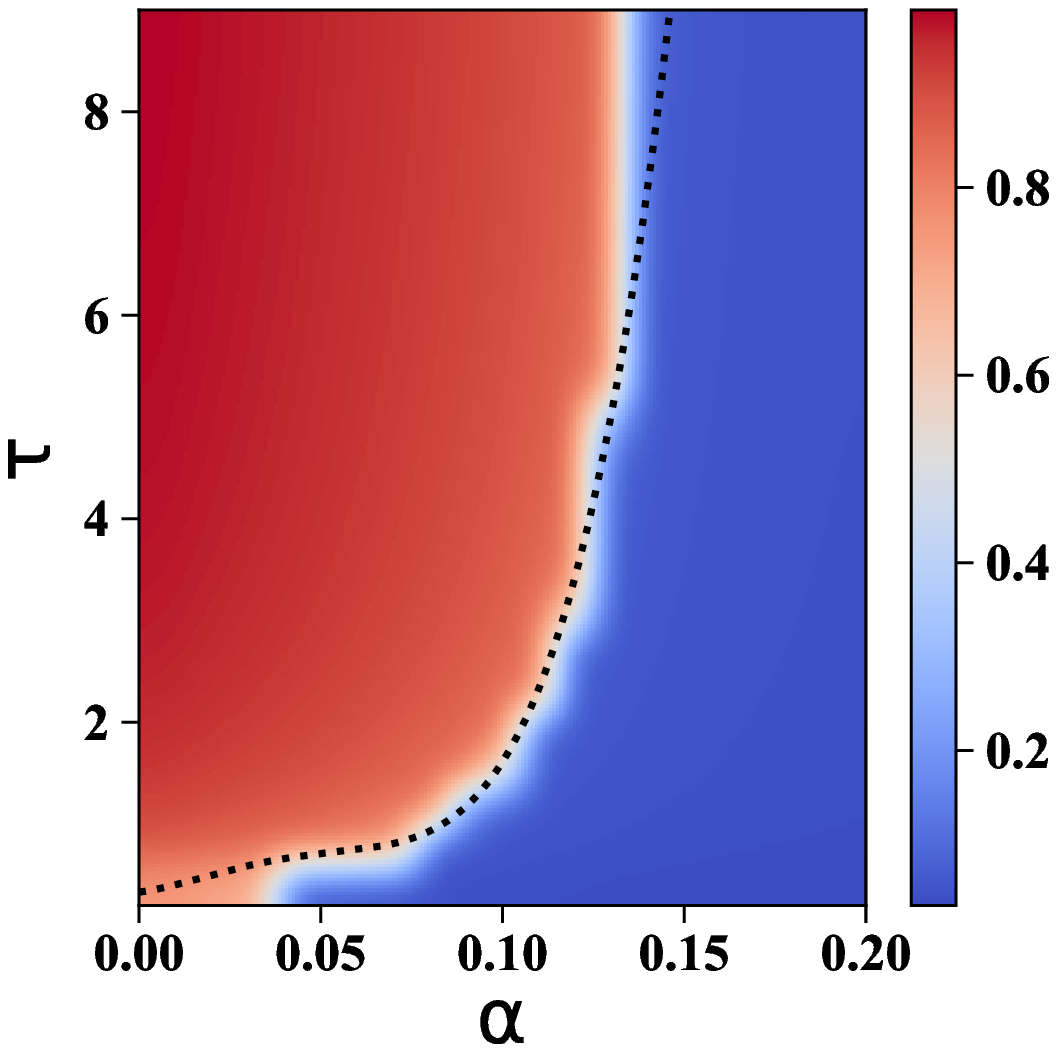}\llap{\parbox[b]{3.8in}{(e)\\\rule{0ex}{2in}}}
    \captionlistentry{}
    \label{NeqPhsPLot_eta_Teff_constant}
    \end{subfigure}
    \begin{subfigure}[t]{0.3\textwidth}
    \centering
    \includegraphics[width = \textwidth]{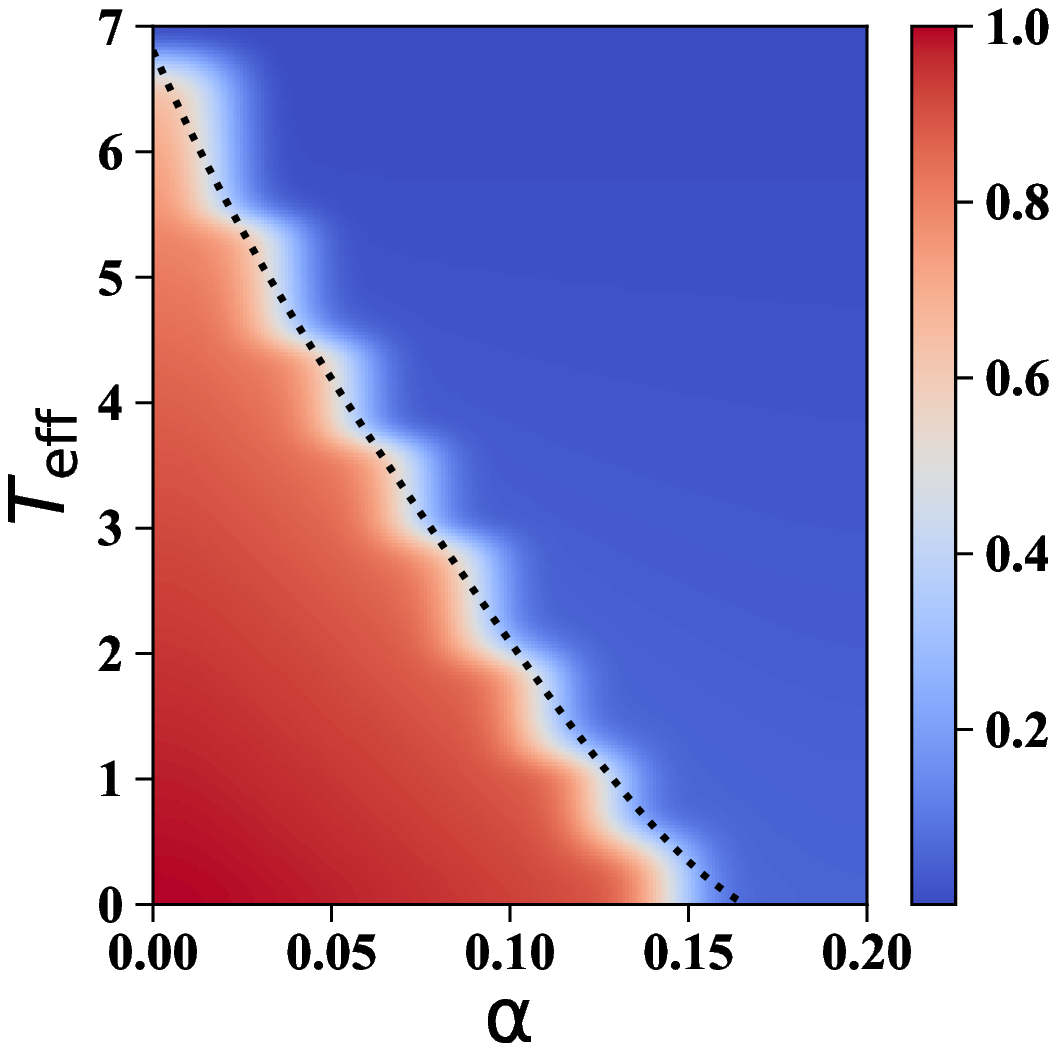}\llap{\parbox[b]{3.8in}{(f)\\\rule{0ex}{2in}}}
    \captionlistentry{}
    \label{NeqPhsPLot_tau_eta_constant}
    \end{subfigure}
    \caption{Enhanced retrieval region from MSR calculations (a) Evolution of $m$ as a function of $t$ from both the full numerical simulations (Eq.~\ref{GLE_SHF_Comparison}) and from the MSR formalism (SI Eqns. \ref{EqnForm_analytical}, \ref{EqnForC_analytical}, \ref{EqnForG_analytical}). Here the solid lines denote the simulation data from 200 spins and the dashed line denotes the analytical prediction. (b) Evolution of $m$ for various passive temperature fractions, $f$, as a function of time using SI Eqns.~\ref{EqnForm_analytical}, \ref{EqnForC_analytical}, \ref{EqnForG_analytical}. As can be clearly seen, with decrease in the passive fraction, retrieval is possible at the same $\alpha$ and $T_{\rm eff}$. (c)-(f) Retrieval is improved when the amount of activity is increased by tuning the persistence time $\tau$, the effective temperature $T_{\rm eff}$ and passive fraction $f$ in the system. $\tau$ determines the time scale over which active fluctuations operate and $f$ controls the ratio of thermal fluctuations to the total fluctuations (thermal + active) in the system. These plots show how the overlap parameter m varies for a range of different parameters. The \textit{blue} region denotes the region where the overlap is very low and there is no retrieval and the \textit{red} region denotes where retrieval is possible. The parameters held fixed for the phase plots are, (c): $\tau = 1$, $T_{\rm eff} = 1$, (d): $\tau = 1$, $\alpha = 0.08$, (e): $T_{\rm eff} = 0.8$, $f = 0$, (f): $\tau = 5.0$, $f = 0$. Here $\tau$ denotes the persistence time of the active noise, $f$ denotes the fraction of passive temperature, $\alpha$ denotes the fraction of stored patterns in the system and $T_{eff}$ denotes the total effective temperature.}
    \label{PhaseDiagramsAndAll}
\end{figure}

\end{widetext}

These equations can be simulated numerically and memory retrieval phase diagrams can be constructed based on the steady state values of m. In Fig. \ref{GLE_SHF_Comparison}, we compare estimates of $m(t)$ obtained from these mean field equations with estimates from full numerical simulations using Eq.~\ref{RelaxationalLangevin} with $N=200$ spins. As can be clearly seen, the mean field equations capture the evolution seen in the full simulations. Further, Fig. \ref{GLE_SHF_Comparison} also demonstrates how the introduction of non-equilibrium activity improve memory recall. Finally, in Fig. \ref{mEvolution} we plot the evolution of $m$ as a function of time, for various values of the fraction $f=\frac{T_p}{T_{\rm eff}}$ (at fixed $\alpha$, $T_{\rm eff}$, and $\tau$). These trends again clearly demonstrate how non-equilibrium forcing improves memory recall. For a fixed $T_{\rm eff}$, memory can be markedly improved if the fluctuations are mainly due to detailed balance violating noise. 

To more comprehensively characterize the improved memory recall due to non-equilibrium dynamics in our system, we use the MSR mean field framework to obtain steady state values of the memory order parameter $m$ at various values of $T_{\rm eff}$, $f$, $\alpha$ and $\tau$ in Fig.~\ref{PhaseDiagramsAndAll}. 
At fixed $\tau$ and $T_{\rm eff}$, increasing the fraction of active temperature increases the capacity as shown in Fig. \ref{NeqPhsPLot_tau_Teff_constant}. At constant $\tau$ and $\alpha$, the system can retrieve better at higher effective temperatures as the fraction $f=T_p/T_{\rm eff}$ is decreased as shown in Fig. \ref{NeqPhsPLot_tau_alpha_constant}. Finally, at constant $T_{\rm eff}$ and $f$, the capacity of the system increases with increasing the persistence time as shown in Fig.~\ref{NeqPhsPLot_eta_Teff_constant}.

These results from the MSR formalism can be physically understood by considering the following effectively single particle caricature. Following Refs \cite{Nandi7688}, it is reasonable to expect that the short time dynamics of the order parameter $m(t)$ in the vicinity of the memory configuration can be approximated by the dynamics of a single particle, $\tilde{\sigma}$ evolving in a harmonic potential and under the influence of an active noise source, 
\begin{align}
    \partial_t \tilde{\sigma} =  - k \tilde{\sigma} + \eta(t) 
\end{align}
where the spring constant $k$ characterizes the restoring force stabilizing the memory states, and the statistics of the noise $\eta$ is the same as that described in Eq.~\ref{NoiseTotal}.

For this system, the probability distribution that the spin state of the particle is $\tilde{\sigma}$ at long times $t$ given it started at time $t=0$ with the state $\tilde{\sigma}_o$ is given by~\cite{Ghosh2021},
\begin{align}
    \lim_{t \to \infty} P(\tilde{\sigma}|\tilde{\sigma}_o;t) = \frac{1}{Z} exp \left( - \frac{k\tilde{\sigma}^2 (1 + \tau k)}{2T_{eff}(1 + \eta\tau k)}  \right) \ , \ \eta = \frac{T_p}{T_{eff}}
\end{align}
where, $Z =\sqrt{2 \pi \frac{k^\prime}{T_{eff}}}$, with $k^\prime \equiv \frac{k(1+\tau k)}{1 + \eta \tau k}$ is the normalization constant and $T_{eff} = T_p + T_a$. This shows that at the same effective temperature, a greater fraction of active temperature leads to an effectively higher spring constant, $k^\prime$ . This is akin to saying that the potential well stabilizing the memory states becomes steeper with activity. This calculation on a minimal model provides intuition for how activity can possibly lead to pattern stabilization and improved associative memory recall.  

Together, these results further show how increasing the non-equilibrium forcing may enhance the associative memory recall of a system. 
\renewcommand\thesubfigure{\alph{subfigure}}
\renewcommand\thesubfigure{\thefigure\alph{subfigure}}
\renewcommand*{\thesection}{\Roman{section}}
\section{Discussion and conclusions}
\label{Conclusions}
\renewcommand*{\thesection}{\arabic{section}}

\begin{figure}[th]
    \begin{subfigure}[t]{0.2\textwidth}
    \centering
    \includegraphics[width =
    \textwidth]{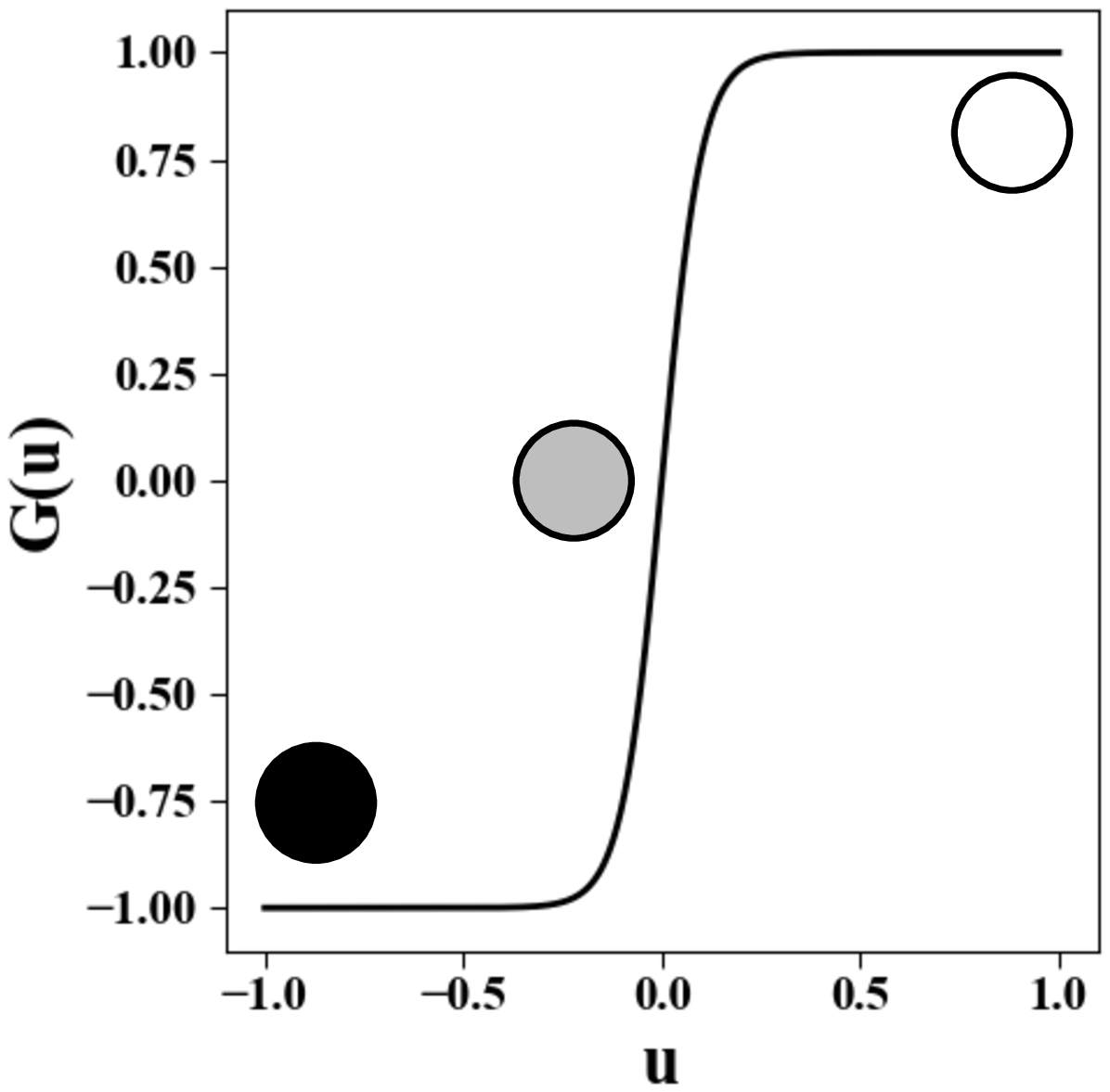}\llap{\parbox[b]{2.7in}{(a)\\\rule{0ex}{1.2in}}}
    \captionlistentry{}
    \label{ActiFunc}
    \end{subfigure}
    \begin{subfigure}[t]{0.2\textwidth}
    \centering
    \includegraphics[width = \textwidth]{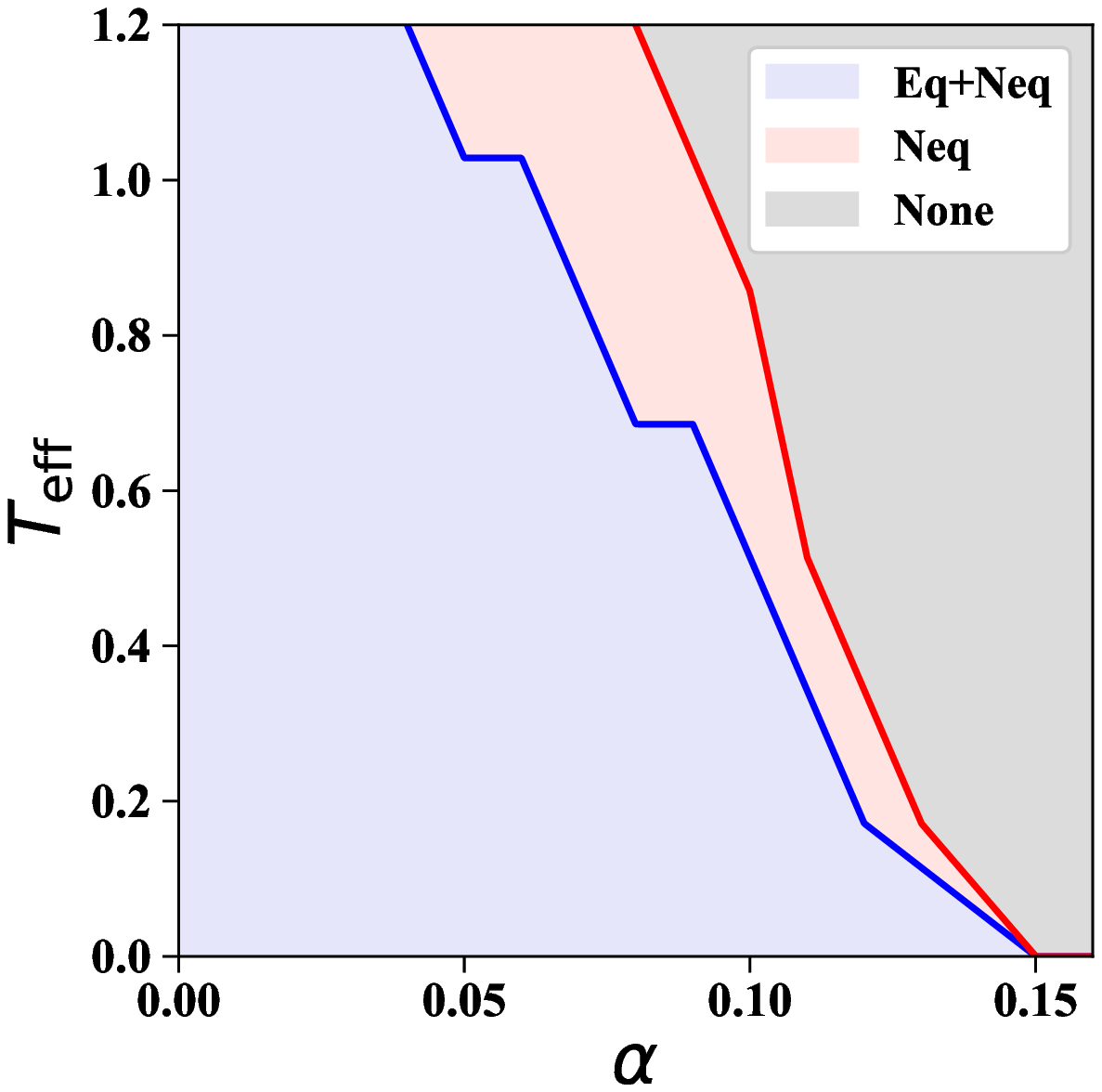}\llap{\parbox[b]{2.7in}{(b)\\\rule{0ex}{1.2in}}}
    \captionlistentry{}
    \label{COntinuous_Spins}
    \end{subfigure}
    \caption{Activity leads to enhanced retrieval region in continuous spins. The colored circles are a schematic to denote the firing of the neurons. In the discrete case, only the \textit{black} (off) and \textit{white} (on) states are available to the neuron. With the continuous activation function, the neurons can access all the other states ranging between 1 and -1 (a) This shows the sigmoid activation function for the neurons. (b) Phase diagram constructed for continuous spin system. Details of the simulation are provided in Supplementary Sec.~\ref{ContNeuronsParams}.}
\end{figure}

Our work here shows that the storage capacity of a system which uses a Hopfield-like strategy to store memory can be increased significantly through the introduction of activity into the system. While we have explored the improved associative memory capacity using specific numerical and analytical tools, it may become possible to develop broader thermodynamic principles for improved associative memory. Indeed, in Fig.~\ref{COntinuous_Spins}, we show that trends uncovered here hold qualitatively for a continuous Hopfield model with neurons firing according to an activation function \cite{Hopfield1984}. 

The equation of motion for the neurons in this case is given by,
\begin{align}
    C_i\frac{du_i}{dt} = J_{ij} G(u_j) - \frac{u_i}{R} + \eta(t)
\end{align}
where ``u" is the state of a given neuron which would correspond to the mean cell potential of the neuron. The input capacitance of the cell membrane is given by C and the transmembrane resistance is denoted by R. C and R determine the charging-discharging timescale of the neuron. They serve the purpose of adding a delay in the response of a given neuron with respect to the firing of neurons around it \cite{Hopfield1984}. $\eta$ has the same statistics as given in Eqns.~\ref{NoiseTotal} and G is the activation function for the neuron and G(u) can be understood as the firing rate of the neuron given the cell potential ``u" (see Fig. \ref{ActiFunc}). The patterns stored in this system would correspond to interneuron connections made when a bunch of neurons fire together. As described in SI Sec.\ref{ContNeuronsParams}, the weight matrix $J_{ij}$ again encodes the stored patterns through a Hebbian like rule. 
Fig.~\ref{COntinuous_Spins} shows how activity in this system improves the retrieval phase. While these qualitative results suggest that non-equilibrium activity may provide a general route to enhance the information processing abilities of a material, we note that there are important caveats. Indeed, the associative memory recall away from equilibrium is dictated by a balance between two competing factors. On the one hand, as our theoretical analysis suggests, non-equilibrium forcing may generate deeper minima in the effective landscape. On the other hand, non-equilibrium forcing also has the potential to generate many spurious minima and hasten the transition to a glassy regime where associative memory properties are lost. Future work will explore these tradeoffs and identify regimes and driving protocols that improve associative memory. 

Finally, following Eq.~\ref{eq:dotwDef}, in the limit of small persistence time, it may become possible to express the change in the effective energetic landscape in terms of the rate of work done by the active forces (SI Sec.~\ref{RateOfWork_supp}). Similar insights have proven useful in the context of active matter systems~\cite{Laura2019,Fodor_2020} to establish connections between dissipation and assembly or organization. Such connections may suggest how non-equilibrium forcing might provide a general mechanism to enhance memory recall \cite{Jeremy2018, Murugan54}.

\section{acknowledgements}
S.V. is supported by funds from DOE BES Grant DE-SC0019765. A.K.B. is supported by funds from ARO under Grant W911NF-14-1-0403. We gratefully acknowledge the helpful discussions we had with Thomas Witten and Sidney Nagel.

\bibliographystyle{unsrt}
\bibliography{Bibliography}

\renewcommand\thesection{\arabic{section}}
\clearpage
\onecolumngrid

\section*{\LARGE Supplementary Information}

\setcounter{figure}{0}
\setcounter{table}{0}
\setcounter{equation}{0}
\setcounter{page}{1}
\setcounter{section}{0}

\renewcommand{\theequation}{S\arabic{equation}} 
\renewcommand{\thepage}{S\arabic{page}} 
\renewcommand{\thesection}{S\arabic{section}}  
\renewcommand{\thetable}{S\arabic{table}}  
\renewcommand{\thefigure}{S\arabic{figure}}

\numberwithin{equation}{section}

\section{Martin-Siggia-Rose approach to dynamics in presence of active noise}
\label{MSRSupp}

For working out the decoupled dynamics of the spins in the Spherical Hopfield model we shall be following the references, Ref. \cite{COOLEN2001619, Bolle2003, Crisanti_dynamics, MSR1973}. The Langevin equation for the spins is given by,
\begin{align}
    \partial_t \sigma_i = -\mu(t) \sigma_i(t) + \theta_i - \frac{\delta \mathcal{H}(\mathbf{\sigma})}{\delta \sigma_i (t)} + \eta_i(t)
\end{align}
where $\mu$ is the Lagrange multiplier that ensures that the spins obey the spherical constraint, $\theta_i$ is the external field at each site, $\mathcal{H}(\mathbf{\sigma})$ is the Hamiltonian and $\eta_i(t)$ is the noise in the system. The noise includes both white noise and active noise.

\begin{align}
    \eta_i(t) &= \eta_{w,i}(t) + \eta_{a, i}(t) \\
    \braket{\eta_{w,i} (t)} &= 0 = \braket{\eta_{a,i} (t)} \ \forall \ i, t \\
    \braket{\eta_{w,i}(t) \eta_{w,j}(t')} &= 2T_p \delta_{ij} \delta(t-t'), \ \braket{\eta_{a,i}(t) \eta_{a,j}(t')} = \frac{T_a}{\tau} \delta_{ij} \exp\left( -\frac{|t-t'|}{\tau} \right), \ \braket{\eta_{w,i}(t) \eta_{a,j}(t')} = 0 \\
    \implies \braket{\eta_i(t) \eta_j(t')} &= 2T_p \delta_{ij} \delta(t-t') + \frac{T_a}{\tau} \delta_{ij} \exp\left( -\frac{|t-t'|}{\tau} \right) = D(t,t')
\end{align}
where we have labelled the entire function as $D(t,t')$. Now we can write the probability of noise, $\eta$ as,
\begin{align}
    P(\eta) \sim \exp\left( -\frac{1}{2} \int dt dt' \eta(t) D^{-1}(t, t') \eta(t') \right) \label{NoiseProbability}
\end{align}

Now we write the generating functional for the system as,
\begin{align}
    Z[\mathbf{\psi}] = \int D\sigma D\eta P(\eta) \exp \left[ \sum_{i=1}^N \int dt \psi_i(t) \sigma_i(t) \right] \prod_{i = 1}^{N} \delta \left(\partial_t \sigma_i(t) + \mu(t)\sigma_i(t) + \frac{\delta \mathcal{H}(\mathbf{\sigma})}{\delta \sigma_i(t)} - \eta_i(t) \right)
\end{align}

The various physical quantities like overlap with a pattern (m), correlations between spins (C) and the response of the spins to an external field (G) can be obtained from differentiating $Z$ w.r.t. the conjugate fields $\psi$ and $\theta$.
\begin{align}
    m^{\mu}(t) &= \sum_{i=1}^N\xi^{\mu}_i \braket{\sigma_i(t)} = -i \lim_{\psi \to 0} \frac{\delta Z[\psi]}{\delta \psi_i(t)} \\
    C_{ij} (t,t') &= \braket{\sigma_1(t) \sigma_j(t')} = \lim_{\psi \to 0} \frac{\delta^2 Z[\psi]}{\delta \psi_i(t) \delta \psi_j(t')} \\
    G_{ij} (t, t') &= \frac{\delta \braket{\sigma_i(t)}}{\delta \theta_j(t')} = -i \lim_{\psi \to 0} \frac{\delta^2 Z[\psi]}{\delta \psi_i(t) \delta \theta_j(t')}
\end{align}

We now express the $\delta$-functions as exponentials and integrate out the noise using the definition of probability given in \eqref{NoiseProbability}. This yields,

\begin{align}
    Z[\psi] &= \int D[\sigma, \hat{\sigma}] \exp \left[ \sum_{i=1}^N \int dt \psi_i(t) \sigma_i(t) + A[\mathbf{\sigma}, \mathbf{\hat{\sigma}}] \right] \\
    A[\mathbf{\sigma}, \mathbf{\hat{\sigma}}] &=  -\frac{1}{2} \sum_{i=1}^{N} \int dt dt' \hat{\sigma_i}(t) D(t,t') \hat{\sigma_i}(t') + i \sum_{i=1}^{N} \int dt \hat{\sigma_i}(t) (L_{0,i} + L_{\xi, i}) \\
    L_{0, i} &= \partial_t \sigma_i(t) + \mu(t) \sigma_i(t) \\
    L_{\xi, i} &= \frac{\delta \mathcal{H}(\mathbf{\sigma})}{\delta \sigma_i} = -\sum_{j}J_{ij} \sigma_j - \sum_{j,k,l} J_{ijkl} \sigma_j \sigma_k \sigma_l = -\frac{1}{N}\sum_{\mu,j, j\neq i}\xi^{\mu}_i \xi^{\mu}_j \sigma_j - \frac{u_0}{N^3}\sum_{\mu,j,k,l} \xi^{\mu}_i \xi^{\mu}_j \xi^{\mu}_k \xi^{\mu}_l \sigma_j \sigma_k \sigma_l
\end{align}

Now we need to take the average of Z over the quenched disorder, $\xi$. Henceforth, $\overline{A}$ would denote the quenched average of A over the pattern variables.

\begin{align}
    \overline{A} &= \int D\xi \exp \left( -\frac{\xi^2}{2} \right) A(\xi) \\
    \overline{Z[\psi]} &= \overline{\int D[\sigma, \hat{\sigma}] \exp \left[ \sum_{i=1}^N \int dt \psi_i(t) \sigma_i(t) + A[\mathbf{\sigma}, \mathbf{\hat{\sigma}}] \right]}
\end{align}
But we observe that the only part in the entire expression that depends on the pattern variables is the part associated with $L_{\xi}$. Thus the quenched average needs to be taken only over $L_{\xi}$. Henceforth, the integrals over time will be implicitly assumed and Einstein summation convention will be used.

We assume that only a single pattern, the first pattern ($\mu=1$) is condensed i.e. the overlap of the spin state with pattern 1 is O(1) at long times and large N. For other patterns it is O(1/N) and decays to 0 for large N. We define a few macroscopic variables,
\begin{align}
    m(t) &= \frac{1}{N} \xi^1_i \sigma_i(t), \ \delta_m = \delta(Nm(t) - \xi^1_i \sigma_i(t)) \\
    w(t) &= \frac{1}{N} \xi^1_i \hat{\sigma_i}(t), \ \delta_w = \delta(Nw(t) - \xi^1_i \hat{\sigma_1}(t)) \\
\end{align}

\begin{align}
    \mathcal{J} &= \overline{\exp\left[ i \hat{\sigma_i} \cdot L_{\xi, i} \right]} = \overline{\exp \left[ -i \hat{\sigma_i} \cdot \left( \frac{1}{N} \xi^{\mu}_i \xi^{\mu}_j \sigma_j + \frac{u_0}{N^3} \xi^{\mu}_i \xi^{\mu}_j \xi^{\mu}_k \xi^{\mu}_l \sigma_j \sigma_k \sigma_l - \frac{1}{N} (\xi^{\mu}_i)^2 \sigma_i  \right) \right]} \label{Lj_average} \\
    &= \overline{\exp \left[ -i N w(t)(m(t) + u_0 m(t)^3) - i  \frac{1}{\sqrt{N}} \xi^{\mu}_i \hat{\sigma_i} \frac{1}{\sqrt{N}} \xi^{\mu}_j \sigma_j + i \frac{1}{N} (\xi^{\mu}_i)^2 \hat{\sigma_i} \sigma_i + O(1/N) \right]} \delta_m \delta_q \label{finalJ}  
\end{align}
Now let us focus on the $3^{rd}$ and $4^{th}$ term inside the exponential because those are the only terms dependent on $\xi$,
\begin{align}
\mathcal{J_2} &= \overline{\exp \left[ - i  \frac{1}{\sqrt{N}} \xi^{\mu}_i \hat{\sigma_i} \frac{1}{\sqrt{N}} \xi^{\mu}_j \sigma_j + i \frac{1}{N} (\xi^{\mu}_i)^2 \hat{\sigma_i} \sigma_i \right]} \\
&= \int D\xi \exp \left[-\frac{1}{2} \sum_{i, \mu}(\xi^{\mu}_i)^2 - i\sum_{\mu} \frac{1}{\sqrt{N}} \sum_i \xi^{\mu}_i \hat{\sigma_i}  \frac{1}{\sqrt{N}} \sum_j \xi^{\mu}_j \sigma_j + i \frac{1}{N} \sum_{\mu, i} (\xi^{\mu}_i)^2 \hat{\sigma_i} \sigma_i \right] \\
&= \int Dx Dy D\xi \exp\left[-\frac{1}{2} \sum_{i, \mu}(\xi^{\mu}_i)^2 \left( 1 - \frac{2i}{N} \sigma_i \hat{\sigma_i} \right) - i \sum_{\mu} x^{\mu} \cdot y^{\mu} \right] \prod_{\mu = 1}^{\alpha N} \left[ \delta(x - \frac{1}{\sqrt{N}} \sum_i \xi^{\mu}_i \sigma_i) \delta(y - \frac{1}{\sqrt{N}} \sum_i \xi^{\mu}_i \hat{\sigma_i}) \right] \\
&= \int Dx Dy D\hat{x} D\hat{y} D\xi \exp[i \sum_{\mu} (x^{\mu}\cdot \hat{x}^{\mu} + i y^{\mu}\cdot \hat{y}^{\mu} - x^{\mu} \cdot y^{\mu})] \nonumber \\
&\exp\left[-\frac{1}{2} \sum_{i, \mu}(\xi^{\mu}_i)^2 \left( 1 - \frac{2i}{N} \sigma_i \hat{\sigma_i} \right) - i\sum_{\mu = 1}^{\alpha N}  \hat{x}^{\mu} \frac{1}{\sqrt{N}} \sum_i \xi^{\mu}_i \sigma_i - i \sum_{\mu=1}^{\alpha N} \hat{y}^{\mu} \frac{1}{\sqrt{N}}  \sum_i \xi^{\mu}_i \hat{\sigma_i} \right] \\
&=  \left[det\left(\mathbb{1} - \frac{2i}{N}\int dt \sum_{i} \hat{\sigma_i}(t) \delta_{ij} \sigma_j(t) \right) \right]^{-\frac{\alpha N}{2}} \nonumber \\
& \left[\exp \left[ -\frac{1}{2N} \int dt dt' \sum_i (\hat{x}(t) \sigma_i(t) + \hat{y}(t) \hat{\sigma_i}(t)) \left( \delta_{ij} - \frac{2i}{N} \int dt \sigma_i(t) \hat{\sigma_i}(t)  \right)^{-1} \sum_i (\hat{x}(t') \sigma_j(t') + \hat{y}(t') \hat{\sigma_j}(t'))  \right] \right]^{\alpha N}
\end{align}

We define a few two-time quantities (which will emerge naturally further down the calculation),
\begin{align}
    K(t,t') &= \frac{1}{N} \sigma_i(t) \hat{\sigma_i}(t') \\
    q(t,t') &= \frac{1}{N} \sigma_i (t) \sigma_i (t') \\
    Q(t,t') &= \frac{1}{N} \hat{\sigma_i} (t) \hat{\sigma_i} (t')
\end{align}

Now we use the following approximations and results,
\begin{align}
    \left( \delta_{ij} - \frac{2i}{N} \int dt \sigma_i(t) \hat{\sigma_i}(t) \right)^{-1} &\approx \mathbb{1} + O(1/N) \\
    det\left(\mathbb{1} - \frac{2i}{N}\int dt \hat{\sigma_i}(t) \sigma_i(t) \right) &= det \left( \mathbb{1} - 2i \delta_{t,t'} K(t, t') \right) \ (Weinstein-Aronsazn \ Theorem)
\end{align}

Thus we have,
\begin{align}
    \mathcal{J_2} &= \int DK \left[ det \left( \mathbb{1} - 2i \delta_{t,t'} K(t, t') \right)\right]^{-\frac{\alpha N}{2}} \delta\left( NK(t,t') - \sum_i \sigma_(t) \hat{\sigma_i}(t) \right) \int Dx D\hat{x} Dy D\hat{y} \exp[i\alpha N(x\cdot \hat{x} + y \cdot \hat{y} - x \cdot y)] \nonumber \\
    &\exp \left[ -\frac{\alpha N}{2} \int dt dt' \sum_i (\hat{x}(t) \sigma_i(t) \sigma_i(t') \hat{x}(t') + 2\hat{x}(t) \sigma_i(t) \hat{\sigma_i}(t) \hat{y}(t') + \hat{y}(t) \hat{\sigma_i}(t) \hat{\sigma_i}(t') \hat{y}(t) ) \right] \\
    &= \int DK \left[ det \left( \mathbb{1} - 2i \delta_{t,t'} K(t, t') \right)\right]^{-\frac{\alpha}{2}} \delta\left( NK(t,t') - \sum_i \sigma_(t) \hat{\sigma_i}(t) \right) \int Dx D\hat{x} Dy D\hat{y} \exp[i(x\cdot \hat{x} + y \cdot \hat{y} - x \cdot y)] \nonumber \\
    &\int DQ Dq \exp\left[ -\frac{\alpha N}{2} \int dt dt' (\hat{x}(t) q(t,t') \hat{x}(t') + 2\hat{x}(t)K(t',t) \hat{y}(t') + \hat{y}(t) Q(t,t') \hat{y}(t')) \right] \nonumber \\
    &\delta\left( Nq(t,t') - \sum_i \sigma_i(t) \sigma_i(t') \right) \delta \left( NQ(t, t') - \sum_i \hat{\sigma_i}(t) \hat{\sigma_i}(t') \right) \delta\left( NK(t,t') - \sum_i \sigma_i(t) \hat{\sigma_i}(t') \right)
\end{align}

Now we express all the remaining $\delta$-functions as exponentials and ignore the constant terms involving $2\pi$. We can also integrate $x$ and $y$. This leads to,
\begin{align}
    \mathcal{J_2} &= \int DK D\hat{K} \left[ det \left( \mathbb{1} - 2i \delta_{t,t'} K(t, t') \right)\right]^{-\frac{\alpha N}{2}} \delta\left( NK(t,t') - \sum_i \sigma_(t) \hat{\sigma_i}(t) \right) \int D\hat{x} D\hat{y} \nonumber \\
    &\int DQ Dq D\hat{Q} D\hat{q} \exp\left[ -\frac{\alpha N}{2} \int dt dt' (\hat{x}(t) q(t,t') \hat{x}(t') + 2\hat{x}(t)K(t',t) \hat{y}(t') + \hat{y}(t) Q(t,t') \hat{y}(t') - 2i \hat{x}(t) \delta(t-t') \hat{y}(t')) \right] \nonumber \\
    &\exp\left[ i\hat{q}(t,t')\left( Nq(t,t') - \sum_i \sigma_i(t) \sigma_i(t') \right) + i\hat{Q}(t,t') \left( NQ(t, t') - \sum_i \hat{\sigma_i}(t) \hat{\sigma_i}(t') \right) + i\hat{K}(t,t')\left( NK(t,t') - \sum_i \sigma_i(t) \hat{\sigma_i}(t') \right) \right] \label{finalJ2}
\end{align}

Putting everything together, \eqref{finalJ} and \eqref{finalJ2} yield,
\begin{align}
    \mathcal{Z[\psi]} &= \int D[\sigma, \hat{\sigma}] DK D\hat{K} Dm D\hat{m} Dw D\hat{w} DQ D\hat{Q} Dq D\hat{q} \exp\left[-\frac{1}{2}\sum_{i=1}^N \int dt dt' \hat{\sigma_1}(t) D(t,t') \sigma_i(t') + i\sum_{i=1}^N \int dt \hat{\sigma_i}(t) L_{0,i} \right] \nonumber \\
    &\exp[N(\Psi(\mathbf{m}, \mathbf{\hat{m}}, \mathbf{w}, \mathbf{\hat{w}}, \mathbf{q}, \mathbf{\hat{q}}, \mathbf{Q}, \mathbf{\hat{Q}}, \mathbf{K}, \mathbf{\hat{K}}) + \Phi(\mathbf{m}, \mathbf{w}, \mathbf{q}, \mathbf{Q}, \mathbf{K}) + \Theta(\mathbf{\hat{m}}, \mathbf{\hat{w}}, \mathbf{\hat{q}}, \mathbf{\hat{Q}}, \mathbf{\hat{K}}))] \label{EqnBeforeSaddlePoint}
\end{align}
where $\Psi$, $\Phi$ and $\Theta$ are defined as follows,
\begin{align}
    \Psi &= i[m(t)\hat{m}(t) + w(t)\hat{w}(t) + q(t,t')\hat{q}(t,t') + Q(t,t')\hat{Q}(t,t') + K(t,t')\hat{K}(t,t')] \\
    \Phi &= -i w(t)[m(t) + u_0 m(t)^3] - \frac{\alpha}{2} \ln \left[ det(1 - 2i\delta_{t,t'} K(t,t')) \right] + \nonumber \\
    &\alpha \ln \left( \int D\hat{x} D\hat{y} \exp\left[ -\frac{1}{2} \int dt dt' (\hat{x}(t) q(t,t') \hat{x}(t') + 2\hat{x}(t)K(t',t) \hat{y}(t') + \hat{y}(t) Q(t,t') \hat{y}(t') - 2i \hat{x}(t) \delta(t-t') \hat{y}(t')) \right] \right) \\
    \Theta &= -i\frac{1}{N} \left[\hat{m}(t)\xi^{1}_i \sigma_i(t) + \hat{w}(t)\xi^1_i \hat{\sigma_1}(t) + \hat{q}(t,t')\sigma_i(t) \sigma_(t') - i\hat{Q}(t,t') \hat{\sigma_i}(t)\hat{\sigma_i}(t') + \hat{K}(t,t')\sigma_i(t)\hat{\sigma_i}(t') -\psi_i(t) \sigma_i(t) \right]
\end{align}

\eqref{EqnBeforeSaddlePoint} is of the form where we can use the saddle point approximation to calculate the integral. For the saddle point equations we set $\partial_v (\Psi + \Phi + \Theta) = 0$ where v is one of the macroscopic variables, $m$, $\hat{m}$, .... , $K$, $\hat{K}$. This yields,
\begin{align}
    w(t) &= \hat{m}(t) = 0, \hat{w}(t) = [m(t) + u_0m(t)^3] \\
    K(t, t') &= iG(t', t), \ G(t', t) = \lim_{\psi \longrightarrow 0} \braket{\sigma(t)\hat{\sigma}(t')}_* \\
    q(t,t') &= -\frac{1}{2}\alpha i  \frac{ \int d\hat{x} d\hat{y} \hat{x}(t) \hat{x}(t') \exp(-\frac{1}{2} [\hat{x} q \hat{x} + 2\hat{y} K \hat{x} - 2i \hat{x} \hat{y} + \hat{y} Q \hat{y}] ) }{ \int d\hat{x} d\hat{y} \exp(-\frac{1}{2} [\hat{x} q \hat{x} + 2\hat{y} K \hat{x} - 2i \hat{x} \hat{y} + \hat{y} Q \hat{y}] ) } = 0 \\
    Q(t, t') &= -\frac{1}{2}\alpha i  \frac{ \int d\hat{x} d\hat{y} \hat{y}(t) \hat{y}(t') \exp(-\frac{1}{2} [\hat{x} q \hat{x} + 2\hat{y} K \hat{x} - 2i \hat{x} \hat{y} + \hat{y} Q \hat{y}] ) }{ \int d\hat{x} d\hat{y} \exp(-\frac{1}{2} [\hat{x} q \hat{x} + 2\hat{y} K \hat{x} - 2i \hat{x} \hat{y} + \hat{y} Q \hat{y}] ) } = -\frac{1}{2} \alpha i \left[ (1 - G)^{-1} C (1 - G^{\dagger})^{-1} \right] (t,t') \\
    K(t, t') &= -\alpha i  \frac{ \int d\hat{x} d\hat{y} \hat{y}(t) \hat{x}(t') \exp(-\frac{1}{2} [\hat{x} q \hat{x} + 2\hat{y} K \hat{x} - 2i \hat{x} \hat{y} + \hat{y} Q \hat{y}] ) }{ \int d\hat{x} d\hat{y} \exp(-\frac{1}{2} [\hat{x} q \hat{x} + 2\hat{y} K \hat{x} - 2i \hat{x} \hat{y} + \hat{y} Q \hat{y}] ) } - \alpha (\mathbb{1} - \delta_{t,t'}K(t,t'))^{-1} \delta(t,t') \\ 
    &= \alpha(1 - G)^{-1} (t, t') - \alpha \mathbb{1} = \alpha G(1-G)^{-1} \\
\end{align}

Substituting these results in \eqref{EqnBeforeSaddlePoint}, we obtain the final result,
\begin{align}
    Z[\psi] = \mathbb{K} \int D[\sigma, \hat{\sigma}] \exp \left[ -\frac{1}{2} \hat{\sigma_i} (D + \alpha [(1 - G)^{-1}C(1 - G^{\dagger})^{-1}]) \hat{\sigma_i} + i\hat{\sigma_i}\left( \partial_t \sigma_i + \mu \sigma_i - [m + u_0 m^3]\xi^1_i - \alpha G(\mathbb{1} - G)^{-1} \sigma_i \right) \right]    
\end{align}
where $\mathbb{K}$ is a constant. From here we can write the effective equation of motion of a single spin which is decoupled from all other spins as,
\begin{align}
    \partial_t \sigma_i (t) + \mu(t) \sigma_i (t) &= [m(t) + u_0 m(t)^3]\xi^1_i + \int dt' \alpha G(\mathbb{1} - G)^{-1} (t,t') \sigma_i(t') + \chi_i (t) \label{DecoupledEqn}\\
    \braket{\chi_i(t) \chi_j(t')} &= \delta_{ij}D(t,t') + \delta_{ij}\alpha \left[ (1-G)^{-1} C (1 - G^{\dagger})^{-1} \right] (t,t') \\
    D(t,t') &= 2T_p \delta(t-t') + \frac{T_a}{\tau} \exp\left(-\frac{|t-t'|}{\tau} \right)\\
    C(t, t') &= \braket{\sigma(t) \sigma(t')}_* \\
    G(t, t') &= \frac{\partial \braket{\sigma(t)}}{\partial \theta(t')}
\end{align}

Using this we can write the equations for m, C and G as follows,

\begin{align}
    \left( \frac{\partial}{\partial t} + \mu(t) \right) m(t) &= [m(t) + u m(t)^3] + \int_{-\infty}^{t} dt' R(t, t') m(t') \label{EqnForm_analytical}\\
    \left( \frac{\partial}{\partial t} + \mu(t) \right) G(t,t') &= \delta(t-t') + \alpha \int_{t'}^t dt_1 R(t, t_1) G(t_1, t') \label{EqnForG_analytical}\\
    \left( \frac{\partial}{\partial t} + \mu(t) \right) C(t,t') &= [m(t) + um(t)^3]m(t') + \alpha \int_{-\infty}^t dt_1 R(t, t_1) C(t', t_1) + 2T_p G(t', t) \nonumber \\
    &+ \alpha \int_{-\infty}^{t'} dt_1 S(t, t_1)G(t', t_1) + \frac{T_a}{\tau} \int_{-\infty}^{t'} dt_1 \exp\left( -\frac{|t-t_1|}{\tau} \right) G(t', t_1) \label{EqnForC_analytical}
\end{align}

\begin{align}
    R(t,t') &= [(1-G)^{-1}G] (t,t') \\
    S(t, t') &= (1-G)^{-1} C (1-G^{\dagger})^{-1} (t,t')
\end{align}

\section{Replica Calculation}
\label{ReplicaCalculationSupp}

In the following calculations we follow Ref. \cite{Dominicis1978, COOLEN2001619}. The Hamiltonian of the Hopfield network is given by,

\begin{align}
    \mathcal{H} (\sigma) =& \frac{1}{2} \mu \sigma_i^2 - \frac{v}{2}J_{ij} \sigma_i \sigma_j - \frac{u}{4}J_{ijkl} \sigma_i \sigma_j \sigma_k \sigma_l \nonumber + \frac{k}{6}J_{ijklmn} \sigma_i \sigma_j \sigma_k \sigma_l \sigma_m \sigma_n + O(1/N) \ terms \\
    v =& 1 + \frac{\tau T_a}{\tilde{T}}(2\mu - 1) \, \ u = u\left[1 + \frac{2\tau T_a}{\tilde{T}}(2\mu - 1) \right] \, \ k = \frac{3\tau T_a}{\tilde{T}} u^2 \\
    J_{ijklmn} =& \frac{1}{N^5} \xi^{\mu}_i \xi^{\mu}_j \xi^{\mu}_k \xi^{\mu}_l \xi^{\mu}_m \xi^{\mu}_n  
\end{align}

The partition function is given by,
\begin{align}
    Z(\beta) = \int_{-\infty}^{\infty}\left[ \prod_{i=1}^{N} d\sigma_i \right] \delta \left( \sum_{i=1}^N  \sigma_i^2 - N \right) \exp[-\beta \mathcal{H}(\sigma)]
\end{align}

We use $\overline{A}$ to denoted the quenched average of macroscopic variable $A$. Now using the replica trick,

\begin{align}
    \beta N f &= -\overline{\ln Z(\beta)} = -\lim_{n \to 0} \frac{1}{n} \overline{Z^n(\beta) - 1} = \lim_{n\to 0}\frac{1}{n}\ln\overline{Z^n} \\
    Z^n(\beta) &= \int_{-\infty}^{\infty} \left[ \prod_{i=1}^N \prod_{\gamma = 1}^n d\sigma_{i}^{\gamma} \right] \prod_{\gamma=1}^n \delta \left( \sum_{i=1}^N (\sigma_i^{\gamma})^2 - N \right) \exp \left[ -\beta \sum_{\gamma=1}^n \mathcal{H}({\sigma_i^{\gamma}}) \right]
\end{align}

where $\gamma$ is the replica index and goes from $1, \ldots , n$. Henceforth, repeated indices imply summation. Only in certain cases will the summation be explicitly denoted.
\begin{align}
    \overline{Z^n(\beta)} =& \int D\sigma D\xi \exp \left[ \beta \left( \frac{v}{2N} \xi^{\mu}_i \xi^{\mu}_j \sigma^{\gamma}_i \sigma^{\gamma}_j + \frac{u}{4N^3} \xi^{\mu}_i \xi^{\mu}_j \xi^{\mu}_k \xi^{\mu}_l \sigma_i^{\gamma} \sigma_j^{\gamma} \sigma_k^{\gamma} \sigma_l^{\gamma} \right. \right. \\
    &- \left. \left. \frac{k}{6N^5} \xi^{\mu}_i \xi^{\mu}_j \xi^{\mu}_k \xi^{\mu}_l \xi^{\mu}_m \xi^{\mu}_n \sigma_i^{\gamma} \sigma_j^{\gamma} \sigma_k^{\gamma} \sigma_l^{\gamma} \sigma_m^{\gamma} \sigma_n^{\gamma} - \frac{v}{2N}\sum_{\mu, i, \gamma} (\xi^{\mu}_i)^2 (\sigma_i^{\gamma})^2 \right) \right] \nonumber \\
    &\prod_{\gamma = 1}^n \delta \left( (\sigma_i^{\gamma})^2 - N \right) \exp\left[ -\frac{(\xi^{\mu}_i)^2}{2} \right]
\end{align}

Now we introduce the overlap parameter $m^{\mu \gamma} = \frac{1}{N} \xi^{\mu}_i \sigma^{\gamma}_i$. For simplicity we assume that only one pattern, pattern number 1, is condensed, i.e. $m^{\mu} \sim O(1)$ for $\mu = 1$ and $O(1/N)$ for $\mu \neq 1$. Using this we can separate the partition function into a condensed and non-condensed part. We introduce this macroscopic variable through a $\delta$ function and express the $\delta$ function as an exponential. Henceforth it is implicitly assumed that m denotes the overlap with only pattern 1. This yields,

\begin{align}
    \int Dm \delta (N m^{\gamma} &- \xi^{1}_i \sigma^{\gamma}_i) = 1 \\
    \int Dm D\tilde{m} \ &\exp\left[ i\tilde{m}^{\gamma} (N m^{\gamma} - \xi^{1}_i \sigma^{\gamma}_i) \right] = 1 \\
    \overline{Z^n(\beta)} &= \int D\sigma D\xi Dm D\tilde{m} \exp(U) \prod \delta((\sigma^{\gamma})^2 - N) \\
    U &= N\beta ( v (m^{\gamma})^2 + u (m^{\gamma})^4 - k (m^{\gamma})^6 ) + \frac{\beta v}{2N} \sum_{\mu \neq 1, i, j, \gamma}\xi^{\mu}_i \xi^{\mu}_j \sigma^{\gamma}_i \sigma^{\gamma}_j - \frac{(\xi^{\mu})^2}{2} + i\tilde{m}^{\gamma}(N m^{\gamma} - \xi^{1}_i \sigma_i) + O(1)
\end{align}

Now we first carry out integration over the quenched disorder, $\xi$. The Relevant terms in U are,
\begin{align}
    -\frac{1}{2}\sum_{\mu \neq 1, i, j} \xi^{\gamma}_i \left( \delta_{ij} - \frac{\beta v}{N} \sum_{\gamma} \sigma^{\gamma}_i \sigma^{\gamma}_j \right) \xi^{\mu}_j - \sum_i \frac{(\xi^1_i)^2}{2} - i\sum_{\gamma}\tilde{m}^{\gamma} \sum_i \xi^1_i \sigma^{\gamma}_i
\end{align}
The $\mu = 1$ integral yields,
\begin{align}
    \exp\left[ -\frac{1}{2} \sum_{\gamma, \kappa} m^{\gamma} \left(\sum_{i} \sigma_i^{\gamma} \sigma_i^{\kappa}\right) m^{\kappa} \right] &= \int DQ \delta\left( NQ^{\gamma \kappa} - \sum_{i} \sigma^{\gamma}_i \sigma^{\kappa}_i \right) \exp \left[ -\frac{N}{2} \sum_{\gamma, \kappa} m^{\gamma} Q^{\gamma \kappa} m^{\kappa} \right] \\
    Q^{\gamma \kappa} &= \delta^{\gamma \kappa} + (1 - \delta^{\gamma \kappa}) q^{\gamma \kappa} \\
    q^{\gamma \kappa} &= \frac{1}{N} \sum_i \sigma_i^{\gamma} \sigma_i^{\kappa} \ \forall \ \gamma \neq \kappa
\end{align}
Thus the correlation of spins between different replicas arises naturally just as in other spin-glass systems.
The $\mu \neq 1$ integral yields,
\begin{align}
    \prod_{\mu = 2}^{\alpha N} \left[ det \left( \delta_{ij} - \frac{\beta v}{N} \sum_{\gamma} \sigma_i^{\gamma} \sigma_j^{\gamma} \right) \right]^{-\frac{1}{2}} = \prod_{\mu = 2}^{\alpha N} \left[ det\left( \delta_{\gamma \kappa} - \frac{\beta v}{N} \sum_{i} \sigma_i^{\gamma} \sigma_i^{\kappa} \right) \right]^{-\frac{1}{2}} = [det(\mathbbm{1} - \beta v \mathbf{Q})]^{-\frac{\alpha N}{2}}
\end{align} 
This step can be carried out through the Weinstein-Aronsazn theorem.
After all this, we have,
\begin{align}
    \overline{Z^n(\beta)} &= \int D\sigma Dm D\tilde{m} Dq D\tilde{q} D\tilde{\lambda} [det(\mathbbm{1} - \beta v \mathbf{Q})]^{-\frac{\alpha N}{2}} e^{U} \\
    U &= N\beta[\frac{v(m^{\gamma})^2}{2} + \frac{u(m^{\gamma})^4}{4} + \frac{k(m^{\gamma})^6}{6}] - \frac{N}{2} \tilde{m}^{\gamma} Q^{\gamma \kappa} \tilde{m}^{\kappa} + iN m^{\gamma} \tilde{m}^{\gamma} + \sum_{\gamma, \kappa, \gamma\neq\kappa} \tilde{q}^{\gamma \kappa} (Nq^{\gamma \kappa} - \sigma^{\gamma}_i \sigma^{\kappa}_i) + \sum_{\gamma} \tilde{\lambda}^{\gamma} (N - (\sigma^{\gamma}_i)^2)
\end{align}
Here we have expressed the $\delta$ functions for the constraints and that for $q$ as exponentials.
Now we integrate over $\tilde{m}$. The relevant terms are,
\begin{align}
    - \frac{N}{2} \tilde{m}^{\gamma} Q^{\gamma \kappa} \tilde{m}^{\kappa} + iN m^{\gamma} \tilde{m}^{\gamma}
\end{align}
Integration yields,
\begin{align}
    det(Q)^{-\frac{1}{2}} \exp \left( -\frac{1}{2}m^{\gamma} Q^{\gamma \kappa} m^{\kappa} \right)
\end{align}
Finally we carry out integration over the spin variables $\sigma$. The relevant terms are,
\begin{align}
    &-\sum_{\gamma, \kappa, i} \sigma^{\gamma}_i \tilde{Q}^{\gamma \kappa} \sigma^{\kappa}_i \\
    \tilde{Q}^{\gamma \kappa} &= \tilde{\lambda}^{\gamma} \delta^{\gamma \kappa} + (1 - \delta^{\gamma \kappa})\tilde{q}^{\gamma \kappa}
\end{align}
Integration yields,
\begin{align}
    [det\tilde{Q}]^{-\frac{N}{2}}
\end{align}

Combining everything, we have,
\begin{align}
    \beta f &=  \lim_{N \to \infty} \lim_{n \to 0} \frac{-1}{n} \ln \frac{1}{N} \left[ \int_{-\infty}^{\infty} Dm \int_{-\infty}^{\infty} Dq \int_{-i\infty}^{i\infty} D\tilde{\lambda} \int_{-i\infty}^{i\infty} D\hat{q} e^{-Ng} \right] \\
    Dm &= \prod_{\gamma=1}^n dm_{\gamma}, \ Dq = \prod_{\gamma \neq \kappa} dq_{\gamma \kappa}, \ D\tilde{\lambda} = \prod_{\gamma=1}^n d\tilde{\lambda}_{\gamma}, \ D\hat{q} = \prod_{\gamma \neq \kappa} d\hat{q}_{\gamma \neq \kappa} \nonumber \\
    g &= -\frac{\beta v}{2} \sum_{\gamma=1}^n m_{\gamma}^2 - \frac{\beta u }{4} \sum_{\gamma=1}^n m_{\gamma}^4 + \frac{\beta k}{6}  \sum_{\gamma=1}^n m_{\gamma}^6 \nonumber + \frac{1}{2} \sum_{\gamma, \kappa=1}^n m_{\alpha} q^{-1}_{\gamma \kappa} m_{\kappa} + \frac{\alpha}{2} ln \ det(\mathbbm{1} - \beta \mathbf{Q}) \nonumber \\
    &+ \frac{1}{2} ln \ det \mathbf{\tilde{Q}} - \frac{1}{2} \sum_{\gamma, \kappa = 1}^n \tilde{Q}_{\gamma \kappa} Q_{\gamma, \kappa} \label{ExpressionFor_g}
\end{align}

This integral can be evaluated through the saddle point method. First, finding the saddle point with respect to $\tilde{Q}$ gives us,
\begin{align}
    \tilde{Q}^{-1}_{ij} = Q_{ij} \implies \tilde{\mathbf{Q}} = \mathbf{Q}^{-1}
\end{align}

Extremizing eith respect to the other variables will be done later. Now we need to assume a form for the matrix $\mathbf{Q}$ and for $m^{\gamma}$. Let us assume that $Q$ is Replica symmetric (RS) i.e. $Q^{\gamma \kappa} = q + (1 - q)\delta^{\gamma \kappa}$ and  $m^{\gamma} = m \ \forall \ \gamma$. We will find the free energy for this form of $Q$ and check for it's stability against RS-breaking fluctuations. As it turns out, for our purposes, the RS form is stable for the retrieval and the paramagnetic phase. It becomes unstable only at extremely low temperatures. 

\begin{align}
    Q_{\gamma \kappa} &= q + (1 - q)\delta_{\gamma \kappa} \\
    Q^{-1}_{\gamma \kappa} &= A\delta_{\gamma \kappa} + B  \, \ A = \frac{1}{1-q} \, \ B = -\frac{q}{(1-q)[]1 + (n-1)q]} = -\frac{q}{(1-q)^2} \ (when \ \lim_{n \to 0}) \\
    det(Q) &= \left( 1 + n \frac{q}{1 - q} \right) det((1-q)\mathbbm{1}) \\
    \ln det(Q) &=  \ln\left[ 1 + n\frac{q}{1 - q} \right] + n\ln(1-q) \\
    \lim_{n\to 0} \frac{1}{n}\ln det(Q) &= \frac{q}{1-q} + \ln(1 - q) \\
    Similarly, \ &\lim_{n\to 0} \frac{1}{n} \ln det(1 - \beta v Q) = \frac{-\beta q v}{1 - \beta v(1 - q)} + \ln(1 - \beta v(1 - q))
\end{align}

All this was found using the Matrix Determinant Lemma. Putting all this back in the expression for f, we finally obtain,

\begin{align}
    \beta f &= extremum\left(-\frac{\beta v}{2} m^2 - \frac{\beta u}{4} m^4 + \frac{\beta k}{6} m^6 + \frac{\alpha}{2} \left[\ln[1 - \beta v(1-q)] - \frac{\beta v q}{1 - \beta v(1-q)} \right] - \frac{1}{2}\left[ \ln(1-q) + \frac{q - m^2}{1 - q} \right] \right)
\end{align}

The saddle point equations for variation across m and q are thus given by,
\begin{align}
    \frac{\partial f}{\partial m} &= 0 \implies m\left[1 + um^2 - km^4 - \frac{1}{\chi} \right] = 0 \\
    \frac{\partial f}{\partial q} &= 0 \implies \frac{\alpha q}{(1 - \chi)^2} = \frac{q - m^2}{\chi^2} \\
    \chi &= \beta(1 - q)
\end{align}

We can investigate the stability of the RS solution by adding RS breaking fluctuations to $Q$. We denote the RS broken matrix as $Q^B$
\begin{align}
    Q^B_{ij} = Q_{ij} + \eta_{ij}, \ \eta_{ij} = \eta_{ji}, \ \eta_{ii} = 0 \\
    (Q^B)^{-1} = [Q + \eta]^{-1} = Q^{-1} - Q^{-1} \eta Q^{-1} + Q^{-1} \eta Q^{-1} \eta Q^{-1}
\end{align}

We ignore terms linear in $\eta$ as the first order variation is set to 0 while taking extrema.
\begin{align}
    f(m, \tilde{Q}^B, Q^B) - f(m, \tilde{Q}, Q) &= T_1 + T_2 + T_3 + T_4 \\
    T_1 &= \sum_{\gamma, \kappa} m^{\gamma} [(Q^B)^{\gamma \kappa} - Q^{\gamma \kappa}] m^{\kappa}  \\
    T_2 &= \frac{\alpha}{2} \ln \frac{det(\mathbbm{1} - \beta(Q + \eta))}{det(\mathbbm{1} - \beta Q)} \\
    T_3 &= \frac{1}{2} \ln \frac{det Q^B}{det Q} \\
    T_4 &= \frac{1}{2} \sum_{\gamma \kappa} [(Q_{\gamma \kappa} + \eta_{\gamma \kappa})\tilde{Q}_{\gamma \kappa} -Q_{\gamma \kappa} \tilde{Q}_{\gamma \kappa}]
\end{align}
After some algebra, it can be easily shown that,
\begin{align}
    T_1 &= m^2(a+bn)^2 \left[ a\sum_{ikz} \eta_{ik} \eta_{kz} + b(\sum_{ik} \eta_{1k})^2 \right] \\
    T_2 &= -\frac{\alpha}{4} \left[ c^2 \sum_{ij} \eta_{ij}^2 + 2cd \sum_{ijk} \eta_{ij} \eta_{jk} + d^2( \sum_{ij}\eta_{ij})^2\right] \\
    T_3 &= \frac{1}{4} \left[ a^2\sum_{ik}\eta_{ik}^2 + 2ab \sum_{ijk} \eta_{ij} \eta_{jk} + b^2 (\sum_{ij} \eta_{ij})^2 \right] \\
    T_4 &= 0 \\
    a &= \frac{1}{1-q}, \ b = -\frac{q}{(1-q)^2}, \ c = \frac{\beta}{1 - \beta(1-q)}, \ d = \frac{\beta^2 q}{(1- \beta(1-q))^2} \\
    f(m, \tilde{Q}^B, Q^B) - f(m, \tilde{Q}, Q) &= A \sum_{ij} \eta_{ij}^2 + B \sum_{ijk} \eta_{ij} \eta_{jk} + D (\sum_{ij} \eta_{ij})^2 \label{QuadraticForm} \\
    A &= -\frac{1}{4}(\alpha c^2 - a^2) \\
    B &= -\frac{\alpha}{2}cd + \frac{1}{2}ab + am^2(a+bn)^2 \\
    D &= -\frac{1}{4}(\alpha d^2 - b^2) + bm^2 (a + bn)^2
\end{align}

In order for the solution to be stable we need the eigenvalues of the quadratic form \eqref{QuadraticForm} to be positive. The eigenvalues of the quadratic form are eigenvalues of the equation,
\begin{align}
    A\eta_{ij} + \frac{B}{2} \sum_k (\eta_{ik} + \eta_{kj}) + D(\sum_{kl} \eta_{kl}) = \Lambda \eta_{ij}
\end{align}

The eigenvalues for this equation are,
\begin{align}
    \Lambda_1 &= A \\
    \Lambda_2 &= \Lambda_1 + (n-2)\frac{B}{2} = A - B + \frac{nB}{2} \\
    \Lambda_3 &= \Lambda_2 + \frac{nB}{2} + n(n-1)D = A-B + n(B-D) + n^2 D 
\end{align}

Following \cite{Crisant_statics}, the relevant eigenvalue is $\Lambda_1 = A$. $A>0$ gives us the region of stable solutions.

\section{Effective Hamiltonian in presence of Active Noise}
The equations of motion for the spins and the active field are given by,
\label{EffectiveHamiltonianSupp}

\begin{align}
    \Gamma_0^{-1} \frac{\partial \sigma_i}{\partial t} &= -\frac{\delta \mathcal{H}}{\delta \sigma_i} + \chi_i(t) + \xi_i(t) \\
    \tau \frac{\partial \xi_i}{\partial t} &= -\xi_i + \eta_i(t) \\
    \braket{\chi_i(t)} = 0 = \braket{\eta_i(t)}, \ \braket{\chi_i(t) \chi_j(t')} &= 2T \delta_{ij} \delta(t - t'), \ \braket{\eta_i(t) \eta_j(t')} = 2T_a \delta_{ij} \delta(t-t') 
\end{align}
Here, $\Gamma_0^{-1}$ sets the microscopic processing time and for simplicity it is taken to be 1. $\chi(t)$ is the $\delta$ correlated white noise and $\xi(t)$ is the active colored noise and $\tau$ is the persistence time. Thus the Fokker-Planck equation is given by,
\begin{align}
    \frac{\partial \rho}{\partial t} &= \frac{\partial}{\partial \sigma_i} \left( h_i \rho - \chi_i \rho + T \frac{\partial \rho}{\partial \sigma_i} \right) + \frac{1}{\tau} \frac{\partial}{\partial \xi_i} \left( \xi_i \rho + \frac{T_a}{\tau} \frac{\partial \rho}{\partial \xi_i} \right) \label{FPE_main} \\
    h_i &= \frac{\delta \mathcal{H}}{\delta \sigma_i}
\end{align}
It is important to note that here the probability distribution is a function of both the spin degrees and the active degrees of freedom as well. In order to recover the marginal distribution wrt only the spins, we need to integrate out the active degrees of freedom. We do this following the procedure outlined in \cite{Maitra2020} and extend it for the general case.
First let us define a few quantities and their properties,
\begin{align}
    R_{n i_1 i_2 ... i_n} &= \int D\xi \xi_{i_1} \xi_{i_2}... \xi_{i_n} \rho \\
    \int D\xi \xi_{i_1} \xi_{i_2}... \xi_{i_n} \frac{\partial}{\partial \xi_i} (\xi_i \rho) &= - \sum_{j=1}^n \delta_{i, i_j} R_{ni_1 i_2...i_{j-1}i_{j+1}..i_n} \ \forall n\geq 1 \\
    \implies \sum_{i=1}^N \int D\xi \xi_{i_1} \xi_{i_2}... \xi_{i_n} \frac{\partial}{\partial \xi_i} (\xi_i \rho) &= -n R_{ni_1 i_2 ... i_n} \\
    \int D\xi \xi_{i_1} \xi_{i_2}... \xi_{i_n} \frac{\partial^2 \rho}{\partial \xi_i^2} &= \sum_{j=1}^n \delta_{i,i_j} \sum_{j'=1, j'\neq j}^n \delta_{i,i_{j'}} R_{(n-2) \prod_{k\neq j,j'} i_k} \\
    \implies \sum_{i=1}^N \int D\xi \xi_{i_1} \xi_{i_2}... \xi_{i_n} \frac{\partial^2 \rho}{\partial \xi_i^2} &= \sum_{i=1}^N \sum_{j=1}^n \delta_{i,i_j} \sum_{j'=1, j'\neq j}^n \delta_{i,i_{j'}} R_{(n-2) \prod_{k\neq j,j'} i_k}
\end{align}
The properties of R are derived through integration by parts and by setting the boundary term to 0.
Now to ease the readability, we introduce a few more definitions,
\begin{align}
    R_0 &\equiv R_0, \ R_{1i} \equiv R_1, \ R_{2ij} \equiv R_2, \ ... R_{n i_1 i_2 ... i_n} \equiv R_n \\
    \delta_{jk} R_0 &\equiv S_2, \ \delta_{ij} R_{1k} + \delta_{ik} R_{1j} + \delta_{jk} R_{1i} \equiv S_3, \ \delta_{ij}R_{2kl} + 5 \ other \ terms \equiv S_4 ... \\
    h_i R_{n-1} &+ T \partial_i R_{n-1} \equiv J_n
\end{align}
Indices implicitly exist for $R_n$ for $n \geq 1$. Basically $S_n \equiv$ $n\choose2$ terms of the form $\delta_{ij} R_{n-2}$. It is implicitly assumed that $S_n$ and $J_n$ also have n indices.

Now from the Fokker-Planck equation in \eqref{FPE_main} we can write down a hierarchy of equations for $R_k$ where the equation for $R_k$ depends on $R_{k+1}$. We can truncate the hierarchy at any k of our choosing.
\begin{align}
    \partial_i J_1 &- \partial_i R_1 = 0  \label{Eqn_for_R_0}\\
    \partial_i J_2 &- \partial_i R_2 - \frac{R_1}{\tau} = 0 \label{Eqn_for_R_1}\\
    \partial_i J_3 &- \partial_i R_3 - \frac{2R_2}{\tau} + \frac{2 T_a}{\tau^2} S_2 = 0 \label{Eqn_for_R_2}\\
    &\dots \\
    \partial_i J_6 &- \partial_i R_6 - \frac{5R_5}{\tau} + \frac{2T_a}{\tau^2} S_5 = 0 \\
    &\dots \\
    \partial_i J_n &- \partial_i R_n - \frac{(n-1)R_5}{\tau} + \frac{2T_a}{\tau^2} S_{n-1} = 0
\end{align}
Say, we want to truncate at $R_5$, we compare the terms at O($\tau$), i.e. $R_5 = \frac{2T_a}{5\tau} S_5$ and substitute it in the previous equation and continue this upto \eqref{Eqn_for_R_0}. After some algebra, we can write the following,
\begin{align}
    \partial_i \left[(h_i + \tilde{T})R_0 \right] &= 2T_a \sum_{n=1}^{\infty}\frac{(-1)^{n+1} \tau^n}{(n+2)!} \partial^{n+2} S_{n+2} + \sum_{n=1}^{\infty} \frac{(-1)^{n+1} \tau^n}{n!} \partial^{n+1} J_{n+1} \label{InfiniteRecursion}\\
    \tilde{T} &= T + T_a \\
    \partial^n &\equiv \partial_{i_1} \partial_{i_@} ... \partial_{i_n}, \ i_1 \neq i_2 .. \neq i_n \\
    \partial_i \partial_i &\equiv \partial_I^2
\end{align}
Now we can simplify this expression further,
\begin{align}
    \partial^n S_n &\equiv \partial_{i_1} \partial_{i_2} ... \partial_{i_n} S_{ni_1 i_2.. i_n} \\
    &= \binom{n}{2} \partial_{i_1}^2 \partial_{i_3} \partial_{i_4} ... \partial_{i_n} R_{(n-2) i_3 i_4 ... i_n} = \binom{n}{2} \partial_i^2 \partial^{n-2} R_{n-2} \\
    \partial^n J_n &\equiv \partial_{i_1} \partial_{i_2} ... \partial_{i_n} J_{ni_1 i_2.. i_n} \\
    &= \partial_{i_n} \partial_{i_{n-1}} ... \partial_{i_2} \partial_{i_1} \left[ h_{i_1} R_{(n-1)i_2i_3...i_n} + T \partial_{i_1} R_{(n-1)i_2i_3...i_n} \right] \\
    &= \partial^{n-1} \partial_i (h_i R_n) + T \partial^{n-1}\partial_i^2 R_{n-1}
\end{align}
Substituting these results in \eqref{InfiniteRecursion} and after some algebra, we obtain,
\begin{align}
    \partial_i \left[ (h_i + \tilde{T}\partial_i) R_0 \right] = - \sum_{n=1}^{\infty} \frac{(-\tau)^n}{n!}[\tilde{T} \partial_i^2 \partial^n R_n + \partial^n\partial_i(h_i R_n)]
\end{align}
Using this relation we can express $R_0$ as $R_0 = R_0^0 + \tau R_0^1 + \tau^2 R_0^2 + \ldots $, where $R_0^1$ is the first order correction in tau to $R_0$, $R_0^2$ the second order correction and so on. Now we will use the notation, $R_n^{k}$ to denote the $k^{th}$ order correction to $R_n$.
At the zeroth order correction for $R_0$, we have,
\begin{align}
    \partial_i \left[ (h_i + \tilde{T}\partial_i) R_0^0 \right] = 0 \\
    \implies R_0^0 = \exp \left( -\frac{H}{\tilde{T}} \right)
\end{align}

At first order correction, we have,
\begin{align}
    \partial_i \left[ (h_i + \tilde{T}\partial_i) R_0^1 \right] = [\tilde{T} \partial_i^2 \partial_j R_{1j}^0 + \partial_j \partial_i(h_i R_{1j})] - \frac{\tau}{2} [\tilde{T} \partial_i^2 \partial_j \partial_k R_{2jk}^0 + \partial_j \partial_k \partial_i (h_i R_{2jk}^{-1})]
\end{align}

From \eqref{Eqn_for_R_2} we have, $R_{2jk}^{-1} = \frac{T_a}{\tau} S_{2jk} = \frac{T_a}{\tau} \delta_{jk}R_0^0$ and from \eqref{Eqn_for_R_1} we have $R_{1j}^0 = -T_a \partial_k S_{2kj} = -T_a \partial_j R_0^0$. Substituting these in the equation above and after some algebra, we obtain,

\begin{align}
    \partial_i \left[ \exp \left( \frac{H}{\tilde{T}} \right) R_0^1 \right] = \frac{T_a}{\tilde{T}} \left( h_{ijj} -\frac{1}{\tilde{T}}h_{ij}h_j \right) = \frac{T_a}{\tilde{T}} \partial_i \left( h_{jj} - \frac{1}{2\tilde{T}}|h_j|^2 \right)
\end{align}

Thus we can now express $R_0$ including the first correction term and from there we can compute the effective Hamiltonian.
\begin{align}
    R_0 &= R_0^0\left[1 + \frac{\tau T_a}{\tilde{T}} \left( h_{jj} - \frac{1}{2\tilde{T}}|h_j|^2 \right) \right] + O(\tau^2) \\
    R_0 &= \exp\left( -\frac{H}{\tilde{T}} \right) \exp\left( \frac{\tau T_a}{\tilde{T}} \left( h_{jj} - \frac{1}{2\tilde{T}}|h_j|^2 \right) \right) \\
    R_0 &= \exp\left( -\frac{H_{eff}}{\tilde{T}} \right), \ H_{eff} = H - \tau T_a \left( h_{jj} - \frac{1}{2\tilde{T}} |h_j|^2 \right) \label{Eqn_for_Heff_supp}
\end{align}
This is the same expression as derived in \cite{Maitra2020}.

Now we turn back to the Hopfield Hamiltonian and using this expression for the effective Hamiltonian, write down the new terms which arise due to activity.
Contribution from the term $h_{ii}$ are O(1) i.e. they are not extensive as will be shown. Thus they can be ignored.
\begin{align}
    \mathcal{H} &= \frac{1}{2} r \sigma_i^2 - \frac{1}{2} J_{ij} \sigma_i \sigma_j - \frac{u}{4} J_{ijkl} \sigma_i \sigma_j \sigma_k \sigma_l \\
    J_{ij} &= \frac{1}{N} \xi^{\mu}_i \xi^{\mu}_j \\
    J_{ijkl} &= \frac{1}{N^3} \xi^{\mu}_i \xi^{\mu}_j \xi^{\mu}_k \xi^{\mu}_l \\
    h_{ii} &= \sum_{i=1}^N r - \frac{3u}{N^3} \sum_{\mu, i, k, l} \xi^{\mu}_i \xi^{\mu}_i \xi^{\mu}_k \xi^{\mu}_l \sigma_k \sigma_l \\
    &= rN - \frac{3u}{N} \sum_{\mu} \sum_{i=1}^N (\xi^{\mu}_i)^2 (m^{\mu})^2
\end{align}
The first term is a constant so it is irrelevant. For the second term, only one of the pattern is condensed, thus $m^{\mu} = m^{\mu} \delta_{\mu, \nu}$, where $\nu$ is the condensed pattern. Putting this back, we see that it is O(1) thus not extensive.
Now we look at $|h_i|^2$. After some algebra and taking into consideration that only one pattern has condensed, we obtain,
\begin{align}
    |h_{i}|^2 &= (r\sigma_i - J_{ij}\sigma_j - J_{ijkl}\sigma_j\sigma_k\sigma_l)^2 \\
    & = (m^{\nu})^2 \left[ \sum_{j=1}^N (\xi^{\mu}_j)^2 - 2rN \right] + 2u(m^{\nu})^4 \left[ \sum_{i=1}^N (\xi^{\nu})^2 - rN \right] + u^2(m^{\nu})^6 \sum_{i = 1}^N (\xi^{\nu}_i)^2
\end{align}
Now, the pattern variables, $\xi^{\mu}_i$ are iid normal random variables. Thus $\sum_{i = 1}^N (\xi^{\nu}_i)^2$ form a $\chi-squared$ distribution with $N\to \infty$ degrees of freedom which essentially becomes a gaussian distribution.with mean N and variance 2N. Thus $\sum_{i = 1}^N (\xi^{\nu}_i)^2 \sim O(N)$. For simplicity we assume, $\sum_{i = 1}^N (\xi^{\nu}_i)^2 = tN$, where $t\in[0,1]$. Irrespective of the value of t, the system with active noise performs better at higher effective temperatures. 

\section{Rate of Work and Changing Free Energy Landscape}
\label{RateOfWork_supp}
From a mechanical argument the rate of energy dissipation can be defined as the rate of work done by the spins on the environment. It is given as,
\begin{align}
    \mathscr{J} &= \braket{\dot{\sigma} \cdot (\dot{\sigma} - \eta_w)} = \frac{N T_a}{\tau} - \braket{\eta_a \cdot \nabla \mathcal{H}}
\end{align}
$\braket{\eta_a \cdot \nabla \mathcal{H}}$ is the only non-trivial contribution to dissipation and this is what we call the rate of work. This is nothing but the average of the dot product of conservative forces and the nonconservative forces (active noise). It has been shown to be important in leading to collision induced slowing and cluster formation in AOUPs and ABPs \cite{Fodor_2020}.

Here, in the limit of small persistence time, $\tau$, we suspect that a similar principle maybe used to understand the increased associative memory property. Here we provide a brief overview of average rate of work in this context and reserve further calculations for future work.
At equilibrium, the free energy of the system for the condensed patterns is given by, 
\begin{align}
    \mathscr{F} = \lim_{n \to 0} \frac{1}{n} g 
\end{align}
where g is given as in Eq.~\ref{ExpressionFor_g} with terms upto $O(m^4)$. When the system is driven out of equilibrium, in the limit of small $\tau$,  additional terms show up in the effective Hamiltonian \textendash $O(\tau)$ terms  which lead to a larger prefactor infront of the quadratic and quartic interaction terms. This in turn affects g as given in Eq.~\ref{ExpressionFor_g} and thus the free energy. Taking inspiration from the definition of rate of work given in \cite{Sekimoto1998, Laura2019}, we define the quantity,
\begin{align}
    \dot{\tilde{w}} &= \dot{w} + \braket{\mu^2 \sigma^2} - \braket{\mu \sigma \nabla \mathcal{H}} \\
    \dot{w} &= \braket{\eta_a \nabla \mathcal{H}}
\end{align} 

Here $\dot{w}$ measures the average rate of work done by the active field and $- \braket{\mu \sigma \nabla \mathcal{H}}$ measures the work done by the Lagrange force (arising from normalization). The term $\braket{\mu^2 \sigma^2}$ is independent of the pattern variables (which are contained in the Hamiltonian). Thus it acts uniformly across the entire configuration space and raises the energy of every configuration by a constant amount. Thus it gives a trivial contribution to the free energy. So effectively, $\dot{\tilde{w}} - \braket{\mu^2 \sigma^2} = \dot{w} - \braket{\mu \sigma \nabla \mathcal{H}} = \dot{w}_a$, is the quantity of interest.

Substituting this in the expression of g and comparing it with Eq.~\ref{Eqn_for_Heff_supp}, the free energy expression in terms of the ``new" rate of work has deeper energy basins and is given by,
\begin{align}
    g_{new} &= g_{old} + \frac{\beta \tau T_a}{2 \tilde{T}} \dot{w}_a \\
    \mathscr{F}_{new} &= \lim_{n \to 0} \frac{1}{n} g_{new}
\end{align}

Using the procedure outlined in \cite{Laura2019}, it can easily shown that,
\begin{align}
    For \ \partial_t \sigma = \mu \sigma &-\nabla_{\sigma} \mathcal{H} + \eta_w(t) + \eta_a(t) \\
    \braket{\mu \sigma \nabla_{\sigma} \mathcal{H}} - \braket{|\nabla_{\sigma} \mathcal{H}|^2} &= - \braket{\eta_a  \nabla_{\sigma} \mathcal{H}} \\
    \because \nabla^2_{\sigma} \mathcal{H} &\sim O(1)
\end{align}
Here, $\eta_a$ is the active noise, $\eta_w$ is the passive noise and $\mu$ is the Lagrange multiplier ensuring spin normalization.

\begin{align}
    g_{new} &= g_{old} + \frac{\beta \tau T_a}{2 \tilde{T}} \dot{\tilde{w}} \\
    &= g_{old} + g_{trivial} + \frac{\beta \tau T_a}{2 \tilde{T}} (\dot{w} - \braket{\mu \sigma \nabla_{\sigma} \mathcal{H}}) \\
    f_{new} &= \lim_{n \to 0} \frac{1}{n} g_{new}
\end{align}

\section{Continuous Neurons}
\label{ContNeuronsParams}
The value of input capacitance was set to 1, transmembrane resistance was set to 10. the activation function used was, $G(u) = \frac{1 - e^{-20u}}{1 + e^{-20u}}$. For constructing the connection matrix, $J_{ij}$, first the ``raw" pattern variables, $\xi^{\mu}_i$,  were drawn from independent gaussian random distributions (these are of the form of ``input" neuron state variables ``u"). After that they were converted into patterns using the Activation function G, i.e. ``processed" patterns are G($\xi^{\mu}_i$) and the connection matrix was constructed from these activated patterns using Hebbian rule, $J_{ij} = \sum_{\mu} G(\xi^{\mu}_i)G(\xi^{\mu}_j)$.

\end{document}